\pdfoutput=1
\documentclass{article}
\usepackage[utf8]{inputenc}
\usepackage[T1]{fontenc}
\usepackage{lmodern}
\fontencoding{T1}\fontfamily{lmr}\selectfont
\DeclareFontShape{T1}{lmr}{bx}{sc}{<->ssub*cmr/bx/sc}{}
\DeclareFontShape{T1}{lmr}{b}{sc}{<->ssub*cmr/bx/sc}{}
\usepackage{microtype}
\usepackage{graphicx}
\usepackage{booktabs}
\usepackage{amsmath,amssymb,mathtools,amsthm}
\usepackage{multirow}
\usepackage{caption}
\usepackage{subcaption}
\usepackage{algorithm,algorithmic}
\usepackage{xcolor}
\usepackage{url}
\usepackage[numbers,sort]{natbib}
\usepackage{hyperref}
\usepackage{xurl}  %
\usepackage{xspace}
\usepackage{tikz}
\usetikzlibrary{arrows.meta,patterns}
\usepackage{dsfont}
\usepackage{bbm}
\usepackage{fontawesome5}

\definecolor{paperred}{HTML}{A31F34}  
\hypersetup{colorlinks=true,linkcolor=paperred,citecolor=paperred,urlcolor=paperred}

\usepackage{etoolbox}
\makeatletter
\def\Hy@subsectionstring{subsection}
\patchcmd{\hyper@makecurrent}{\ifx\Hy@param\Hy@chapterstring}{%
  \ifx\Hy@param\Hy@subsectionstring
    \ifx\Hy@chapapp\Hy@appendixstring\let\Hy@param\Hy@chapapp\fi
  \fi
  \ifx\Hy@param\Hy@chapterstring}{}{%
  \@latex@warning@no@line{\string\autoref\space prints "Section" for appendix subsections}}
\makeatother
\newcommand{\appref}[1]{\hyperref[#1]{Appendix~\ref*{#1}}}

\newcommand{\instlogo}[2][0.30in]{\raisebox{-0.5\height}{\includegraphics[height=#1]{#2}}}
\newcommand{\logorow}{%
  \makebox[\textwidth][s]{%
    \instlogo[0.32in]{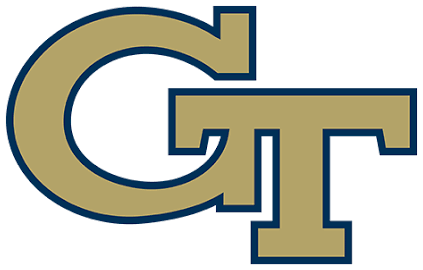}\hfill
    \instlogo[0.17in]{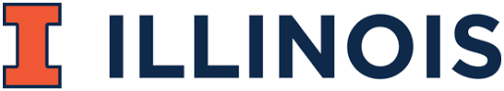}\hfill
    \instlogo[0.18in]{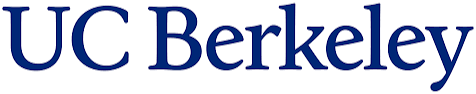}\hfill
    \instlogo[0.21in]{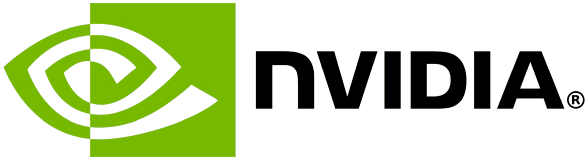}\hfill
    \instlogo[0.21in]{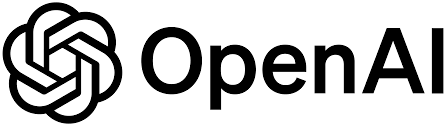}\hfill
    \instlogo[0.29in]{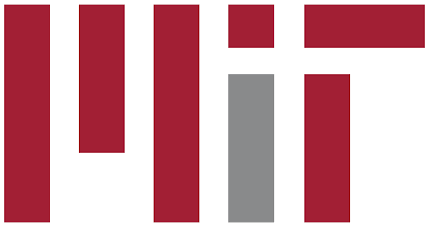}}}
\newcommand{\titleicon}{%
  \smash{\raisebox{\dimexpr0.35em-0.5\height\relax}{%
    \includegraphics[height=1.3em]{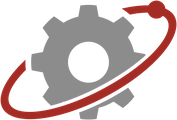}}}\hspace{0.4em}}

\newcommand{\sys}{\textsc{Evolving Infra}}
\newcommand{\papertitle}{Evolving Infra: Agent-Adaptive LLM Serving at Scale}
\hypersetup{pdftitle={\papertitle}}
\makeatletter
\def\@maketitle{%
  \newpage
  \begingroup
    \setlength{\parindent}{0pt}%
    \setlength{\parskip}{0pt}%
    \fontfamily{ppl}\selectfont
    \vspace*{-36pt}%
    \noindent\logorow\par
    \vspace{9pt}%
    \hrule height 0.4pt
    \vspace{15pt}%
    {\fontsize{17}{21}\selectfont\bfseries\raggedright \titleicon\@title\par}%
    \vspace{13pt}%
    {\fontsize{10}{13}\selectfont\raggedright \@author\par}%
    \vspace{6pt}%
    {\fontsize{9.5}{12}\selectfont\raggedright \paperaffiliations\par}%
    \vspace{8pt}%
    {\fontsize{9}{11}\selectfont \paperauthornote\par}%
    \vspace{6pt}%
    {\fontsize{9.5}{12}\selectfont \paperlinks\par}%
  \endgroup
  \vspace{14pt}}
\makeatother

\renewenvironment{abstract}{%
  \par\begingroup
  \fontfamily{ppl}\fontsize{10}{12}\selectfont
  \setlength{\parindent}{0pt}%
  \setlength{\parskip}{0pt}%
  {\fontsize{11}{13}\selectfont\bfseries \abstractname\par}%
  \vspace{5pt}\noindent\ignorespaces
}{%
  \par\endgroup\vspace{6pt}}

\title{\papertitle}

\author{%
  \textbf{Hao~Kang}\textsuperscript{1,*},
  \textbf{Weili~Xu}\textsuperscript{2,*},
  \textbf{Ziyang~Li}\textsuperscript{*},
  \textbf{Zheng~Du}\textsuperscript{1,*},
  \textbf{Hanchen~Li}\textsuperscript{3},
  \textbf{Tian~Xia}\textsuperscript{3},
  \textbf{Song~Bian}\textsuperscript{4},
  \textbf{Tushar~Krishna}\textsuperscript{1},
  \textbf{Huizi~Mao}\textsuperscript{5},
  \textbf{Ion~Stoica}\textsuperscript{3}, and
  \textbf{Song~Han}\textsuperscript{6}}
\newcommand{\paperaffiliations}{%
  \mbox{\textsuperscript{1}Georgia Institute of Technology},
  \mbox{\textsuperscript{2}University of Illinois Urbana-Champaign}\\
  \mbox{\textsuperscript{3}University of California, Berkeley},
  \mbox{\textsuperscript{4}NVIDIA},
  \mbox{\textsuperscript{5}OpenAI},
  \mbox{\textsuperscript{6}Massachusetts Institute of Technology}}
\newcommand{\paperauthornote}{%
  \textsuperscript{*}Equal contribution.\quad
  Corresponding author:
  \href{mailto:hao.kang.sys@gmail.com}{\nolinkurl{hao.kang.sys@gmail.com}}}
\newcommand{\paperlinks}{%
  \href{https://inference-rsi.com/}{\faGlobe\ Webpage}\hspace{1.5em}%
  \href{https://github.com/mit-han-lab/EvolvingInfra}{\faGithub\ Code}}
\date{}

\begin{document}

\maketitle

\begin{abstract}

Agentic workloads are a major source of LLM inference demand.
Existing prefill--decode (PD) disaggregated serving systems offer limited
flexibility for runtime adaptation of compute allocation and KV cache
management, and rely on coarse signals that only partly capture the
diverse and dynamic agent workload. These limitations lead
to load imbalance between prefill and decode nodes and eviction of KV caches that agents will reuse. We present \sys{}, an agent-adaptive serving
framework that adapts compute allocation to agent workloads and KV
cache management to agent context management at runtime. We decouple
compute allocation from key--value (KV) cache
management, allowing nodes to switch efficiently between prefill and decode. Building on this design, an
online planner matches the PD split to workload composition, while an
agent-aware distributed KV manager guides cache eviction and transfer with signals from
agent context management. Across coding, math, and computer-use
tasks with GLM-5.2 and Kimi K2.6 on up to 256 GB300 GPUs, \sys{} reduces
mean end-to-end request latency and P95 TTFT by up to 4.1$\times$ and
14$\times$ over Dynamo, NVIDIA's production framework. We open-source the system
implementation of \sys{} at \url{https://github.com/mit-han-lab/EvolvingInfra}.

\end{abstract}

\begin{center}
  \vspace{-3mm}
  \captionsetup{type=figure}
  \definecolor{eiRed}{HTML}{A31F34}%
\definecolor{eiGray}{HTML}{8C8C8C}%
\definecolor{eiGrayL}{HTML}{CDCDCD}%
\definecolor{eiGrayD}{HTML}{5E5E5E}%
\definecolor{eiInk}{HTML}{1A1A1A}%
\begin{tikzpicture}[
    x=1cm, y=1cm,
    font=\fontfamily{DejaVuSans-TLF}\selectfont\scriptsize,
    >={Stealth[length=3.2pt, width=2.6pt]},
    eititle/.style={font=\fontfamily{DejaVuSans-TLF}\selectfont\fontsize{7}{8.2}\selectfont\bfseries},
    eilegend/.style={anchor=west, inner xsep=0pt,
      font=\fontfamily{DejaVuSans-TLF}\selectfont\fontsize{6}{7}\selectfont},
    box/.style={draw=eiInk, line width=0.6pt, rounded corners=2pt, fill=white},
    gnode/.style={draw=eiInk, line width=0.6pt, rounded corners=2.5pt, fill=white},
    ghost/.style={draw=eiRed, line width=0.9pt, dashed, rounded corners=2.5pt},
    chip/.style={minimum size=0.64cm, inner sep=0pt, line width=0.45pt, rounded corners=1pt,
      anchor=south west, font=\fontfamily{DejaVuSans-TLF}\selectfont\fontsize{6}{7}\selectfont\bfseries},
    chipP/.style={fill=eiInk!82, draw=eiInk, text=white},
    chipD/.style={fill=eiGrayL, draw=eiGrayD, text=eiInk},
    chipG/.style={fill=eiInk!45, draw=eiGrayD, dashed, text=white},
    port/.style={chip, minimum size=0.34cm, anchor=center},
    pin/.style={line width=0.4pt, draw=eiGrayD},
    kv/.style={minimum size=0.29cm, inner sep=0pt, line width=0.4pt, anchor=south west,
      rounded corners=0.6pt,
      font=\fontfamily{DejaVuSans-TLF}\selectfont\fontsize{6}{7}\selectfont\bfseries},
    other/.style={kv, fill=eiGrayL, draw=eiGray, text=eiInk},
    req/.style={kv, fill=eiRed, draw=eiRed!70!black, text=white},
    dead/.style={kv, fill=white, draw=eiGray, densely dashed},
    route/.style={->, line width=0.7pt, draw=eiInk, rounded corners=2pt},
    pull/.style={->, line width=1pt, draw=eiRed, rounded corners=2pt},
    switch/.style={->, line width=1.1pt, draw=eiRed, dash pattern=on 2.6pt off 1.4pt},
  ]
  \def\nb{1.52}\def\nt{2.52}\def\gb{1.7}\def\ch{1.27}
  \def\chips#1#2{%
    \foreach \c in {0.14,0.92} {
      \foreach \t in {0.16,0.32,0.48} \draw[pin]
        (#1+\c+\t,\gb) -- ++(0,-0.05) (#1+\c+\t,\gb+0.64) -- ++(0,0.05)
        (#1+\c,\gb+\t) -- ++(-0.05,0) (#1+\c+0.64,\gb+\t) -- ++(0.05,0);
      \node[chip, #2] at (#1+\c,\gb) {GPU};
    }}
  \def\gpunode#1#2{\draw[gnode] (#1,\nb) rectangle (#1+1.7,\nt); \chips{#1}{#2}}

  \draw[eiGrayD, line width=0.6pt, rounded corners=3pt, fill=black!4] (0.06,0.06) rectangle (10.14,1.12);
  \node[eititle, anchor=base west] at (0.16,0.16) {Shared host memory};
  \node[eilegend, anchor=base east] (deadkey) at (10.04,0.16) {Evicted when full};
  \node[dead, minimum size=0.24cm, anchor=east] (deadbox) at ([xshift=-3pt, yshift=2.3pt]deadkey.base west) {};
  \draw[eiRed, line width=0.6pt] (deadbox.south west) ++(0.03,0.03) -- ++(0.18,0.18)
    (deadbox.north west) ++(0.03,-0.03) -- ++(0.18,-0.18);
  \node[eilegend, anchor=base east] (kvkey) at ([xshift=-0.62cm]deadbox.west |- deadkey.base)
    {KV referenced by \textit{n} branches};
  \node[other, minimum size=0.24cm, anchor=east, font=\fontfamily{DejaVuSans-TLF}\selectfont
    \fontsize{5}{6}\selectfont\bfseries\itshape] at ([xshift=-3pt, yshift=2.3pt]kvkey.base west) {n};
  \foreach \x/\n in {0.4/4, 0.98/4, 2.14/3, 2.72/3, 3.3/2, 3.88/2, 4.46/2, 5.04/2, 6.2/1,
      6.78/1, 7.36/1, 7.94/1}
    \node[other] at (\x,0.5) {\n};
  \node[req] (kvA) at (1.56,0.5) {3};
  \node[req] (kvB) at (5.62,0.5) {2};
  \node[dead, text=eiGray] at (9.1,0.5) {0};
  \node[dead] at (9.68,0.5) {};
  \draw[eiRed, line width=0.8pt] (9.715,0.535) -- (9.935,0.755) (9.715,0.755) -- (9.935,0.535);
  \node[req] (kvN) at (8.52,0.64) {1};

  \gpunode{0.18}{chipP}
  \gpunode{2.06}{chipP}
  \gpunode{6.44}{chipD}
  \gpunode{8.32}{chipD}
  \draw[ghost] (3.94,\nb) rectangle (5.64,\nt);
  \foreach \c in {0.14,0.92} \node[chip, chipG] at (3.94+\c,\gb) {GPU};
  \draw[eiRed, line width=1pt, rounded corners=2.5pt] (6.44,\nb) rectangle (8.14,\nt);
  \draw[switch] (4.79,\nt) to[out=30, in=150] (7.29,\nt);

  \foreach \nm in {kvA, kvB} \draw[pull] (\nm.north) |- (2.91,\ch) -- (2.91,\nb);
  \draw[pull] (8.665,\nb) -- (kvN.north);

  \node[box, minimum width=2.36cm, minimum height=0.52cm, anchor=south west] (rt) at (1.51,3.16) {};
  \node[eititle, anchor=west] at (1.56,3.42) {Router};
  \node[port, chipP] at (2.91,3.42) {P};
  \node[port, chipD] at (3.46,3.42) {D};
  \node[box, eititle, fill=eiRed, draw=eiRed!70!black, text=white, inner xsep=2pt,
    minimum height=0.52cm, anchor=south west] (req) at (0.04,3.16) {Request};
  \draw[route] (req.east) -- (rt.west);
  \draw[route] (2.91,3.16) -- (2.91,\nt);
  \draw[route] (3.46,3.16) -- (3.46,3.02) -- (9.17,3.02) -- (9.17,\nt);

  \node[chip, chipP, minimum size=0.24cm, anchor=west] at (4.52,3.56) {};
  \node[eilegend] at (4.84,3.56) {Prefill resources};
  \node[chip, chipD, minimum size=0.24cm, anchor=west] at (4.52,3.28) {};
  \node[eilegend] at (4.84,3.28) {Decode resources};
  \draw[pull] (7.36,3.56) -- ++(0.42,0);
  \node[eilegend] at (7.88,3.56) {Memory adaptation};
  \draw[switch] (7.36,3.28) -- ++(0.42,0);
  \node[eilegend] at (7.88,3.28) {Compute adaptation};
\end{tikzpicture}\hfill
  \includegraphics[width=0.31\textwidth]{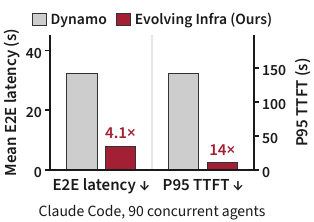}
  \vspace{-3mm}
  \caption{\textbf{Overview and speedup of \sys{}.}
      }
  \label{fig:teaser}
\end{center}

\section{Introduction}
\label{sec:introduction}

Agentic workloads account for a large and growing share of LLM inference~\cite{aubakirova2026state,liu2026agentic}, spanning software engineering~\cite{jimenez2024swebench}, math research~\cite{novikov2025alphaevolve}, and computer use~\cite{anthropic2024computeruse}. Agents execute these tasks in steps, each consisting of a model request and the tool calls it triggers, with harnesses updating the context between steps.
Successive requests therefore typically share long prefixes, creating
opportunities for key--value (KV) cache reuse across consecutive steps.

\begin{figure}[t]
  \centering
  \begin{subfigure}[t]{0.325\textwidth}
    \centering
    \includegraphics[width=\linewidth]{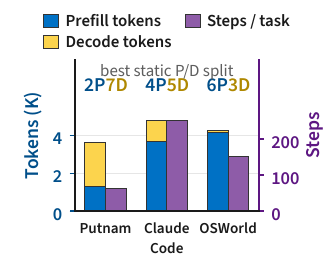}
    \vspace{-7mm}
    \caption{Diverse agent workloads}
    \label{fig:intro-diversity}
  \end{subfigure}\hfill
  \begin{subfigure}[t]{0.325\textwidth}
    \centering
    \includegraphics[width=\linewidth]{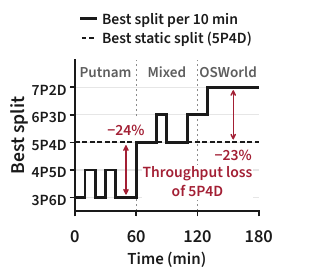}
    \vspace{-7mm}
    \caption{Shifting workload composition}
    \label{fig:intro-dynamicpd}
  \end{subfigure}\hfill
  \begin{subfigure}[t]{0.325\textwidth}
    \centering
    \includegraphics[width=\linewidth]{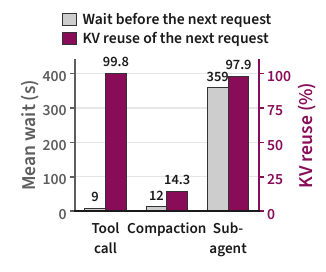}
    \vspace{-7mm}
    \caption{Context-dependent KV reuse}

    \label{fig:intro_context}
  \end{subfigure}

  \vspace{-2.5mm}
  \caption{\textbf{Agentic workloads vary across agents, over time, and within each agent.}
(a) PutnamBench~\citep[Putnam;][]{tsoukalas2024putnambench}, Claude
Code~\citep[CC;][]{semianalysis2026cctraces}, and OSWorld
2.0~\citep[OSWorld;][]{yuan2026osworld} agents differ in prefill and decode work
per request, so their best static splits range from 2P7D to 6P3D ($x$P$y$D: $x$
prefill and $y$ decode nodes). (b) As the workload composition shifts from Putnam
to OSWorld, the best split per 10-minute window moves from 3P6D to 7P2D, and the
best static split (5P4D) loses up to 24\% of throughput in the Putnam and OSWorld
periods. (c) In CC, 98\% of the prompt after a 359\,s sub-agent call matches
earlier KV, but only 14\% after a 12\,s compaction, so recency does not predict KV
reuse.}
  \label{fig:intro}
  \vspace{-3mm}
\end{figure}

These workloads place diverse demands on prefill and decode resources. Differences in tasks, harnesses, and models lead to different amounts of uncached input and generated output per request, making some agent workloads prefill-heavy and others decode-heavy (\autoref{fig:intro-diversity}). The agent workload composition, i.e., the proportions of different agent workloads served by a cluster, also varies over time. A production study of 11.7 million requests finds that user-triggered task starts follow workday rhythms, whereas automatically triggered task starts are more evenly distributed throughout the day~\cite{zheng2026semanticsworkflowsinfrastructureunderstanding}. In prefill--decode (PD) disaggregation, the allocation of nodes (the PD split) that best balances prefill and decode can change as this composition shifts (\autoref{fig:intro-dynamicpd}).

Meanwhile, within an agent, context management changes which prefixes
subsequent requests can reuse. Harnesses compact histories, fork and create
sub-agents~\cite{anthropic2026subagents,openai2025codexmax}. Each fork or sub-agent starts a new \emph{branch}, a context that has its own
sequence of requests and remains \emph{active} until it finishes. A branch waiting for a tool or sub-agent may reuse its existing
prefix after a long pause, whereas compaction can remove recently accessed
history from the subsequent requests of a branch
(\autoref{fig:intro_context}). Access recency alone thus does not
reveal which cached prefixes remain useful to active branches.

Serving agents efficiently thus requires adapting compute allocation to diverse agent workloads and their shifting composition, and KV cache management to agent context management. Production
systems~\cite{nvidia2025dynamo} serve agentic workloads with PD disaggregation~\cite{zhong2024distserve}
and manage compute and memory with three components. A planner determines the compute allocation between prefill and decode by assigning nodes to each phase~\cite{nvidia2026planner,wu2025arrow}. A cache-aware router routes each request to a node
with matching prefix KV~\cite{sglang2026sglang,nvidia2026routing}. A memory pool shares and evicts KV across nodes~\cite{qin2025mooncake,xie2025sglang}. %
However, agentic workloads expose three problems in existing systems.

\textbf{Costly phase switches.} Each adaptation of the compute allocation
switches nodes between phases, and such switches are expensive in existing
systems. Some~\cite{nvidia2026planner} relaunch serving workers to change their phase, which takes minutes and removes the node from service during the switch. Others switch without a restart~\cite{wu2025arrow} but strand the node's cached KV, forcing later requests to recompute their prefixes. In our measurements on Claude Code traces, Dynamo takes 681~s per switch on average, while Arrow loses 55.5\% of the available KV reuse over 34 in-place switches (\autoref{fig:motiv-cost}).

\textbf{Misestimated compute demand.} A PD split profiled offline for each agent
workload is costly to obtain and cannot follow the shifting workload composition.
Even the best static split loses up to 24\% of throughput in some periods
relative to the best split of each 10-minute window (\autoref{fig:intro-dynamicpd}).
Online planners also misestimate demand. They adjust the PD split using coarse
signals such as queue depth~\cite{nvidia2026planner} or predicted time to first
token (TTFT)~\cite{liu2025xllm}. Context management, such as spawning several
sub-agents at once, can create short bursts of prefill work that perturb these
signals and trigger costly phase switches without sustained throughput gains.
Some planners also estimate prefill work from the full input
length~\cite{wu2025arrow}, ignoring KV reuse and overestimating the number of
prefill nodes needed (\autoref{fig:motiv-compute}).

\textbf{Misjudged KV reuse.} Existing systems misjudge KV reuse when they route requests and evict KV. Requests of the same branch share long prefixes. Cache-aware routers~\cite{sglang2026sglang} concentrate these requests on the few prefill nodes holding their prefixes, overloading them and raising P95 TTFT. Routers~\cite{nvidia2026routing} that trade cache locality for load balance can route requests to nodes that must recompute their prefixes. The memory pool avoids this recomputation, but its KV fetches can delay prefill and leave duplicate copies in the pool (\autoref{fig:motiv-memory}). The memory pool and node caches typically evict KV in least recently used (LRU)
order~\cite{zheng2024sglang,qin2025mooncake}. They thus drop the KV of branches
waiting for tools, sub-agents, or users while keeping the history replaced by
compaction, the contexts of finished sub-agents, and duplicate copies left by KV
transfers.

Two limitations underlie these problems. First, compute allocation and KV cache management remain coupled. Engines index cached KV in local state. Switching a node from prefill to decode thus discards or strands its cached KV. Shared KV pools
avoid this coupling, but keep the allocation fixed or change it by relaunching nodes. Second, the three components rely on coarse signals,
such as request latency, input lengths, and access history, which do not
fully reveal the agent workload or context management. To our knowledge, \sys{} is the first agent-adaptive serving framework, adapting compute allocation to diverse agent workloads and their shifting composition, and KV cache management to agent context management at runtime. We make three contributions.

\textbf{Decoupled compute allocation and KV cache management.} To reduce the cost
of phase switches, \sys{} separates the phase of each node from the KV storage.
Each node supports both phases, and the planner reassigns its phase at runtime
without a restart. KV resides in a shared pool formed by the host memory of all
nodes, and a cluster-wide KV directory records the number of active branches that
reference each copy. Neither the pool nor the directory is tied to node phases,
and a phase switch leaves the KV readable by all nodes. The directory updates
these counts in response to agent context management and KV transfers between
nodes~(\autoref{sec:dataplane}).

\textbf{Agent-adaptive compute allocation.} To precisely size the PD split for the
agent workload composition, \sys{} learns iteration-time models of prefill and
decode online, without offline profiling (\autoref{sec:design-model}). The
prefill model incorporates KV reuse, enabling accurate estimation of prefill time
on agentic workloads. The planner evaluates these models on the average work of a
request over each planning window. These averages characterize the workload
itself rather than the momentary load of the cluster. The planner switches only
when the predicted throughput gain of a new split persists across the window,
which keeps short bursts from triggering switches (\autoref{sec:design-split}).

\textbf{Agent-adaptive KV cache management.} To keep the reusable KV of active
branches while balancing the load of prefill nodes, \sys{} routes requests and
evicts KV according to agent context management. It fetches remote KV
asynchronously while a request is queued, which allows the router to send each
prefill request to the node with the shortest predicted TTFT. The directory shows
that no active branch references replaced histories, the contexts of finished
sub-agents, or duplicate copies left by transfers, and \sys{} evicts this KV
first~(Sections~\ref{sec:dataplane-pool}, \ref{sec:dataplane-directory},
and~\ref{sec:design-placement}).

We evaluate \sys{} on coding, computer-use and math agents, serving GLM-5.2 and
Kimi K2.6 on 36 GB300 GPUs. When serving a single agent workload at a
\textbf{stationary load}, \sys{} achieves up to 2.7$\times$ the throughput of
Dynamo and up to 38\% more than the best static split. It reduces the mean
end-to-end (E2E) request latency by up to 4.1$\times$ and P95 TTFT by up to
14$\times$ relative to Dynamo (\autoref{fig:teaser}). For a \textbf{shifting
workload composition} from math to computer-use agents, \sys{} adapts the PD split
online, achieving 2.1$\times$ the throughput of Dynamo and 28\% more than the best
static split (\autoref{sec:eval-shift}).

\section{Background}
\label{sec:background}

\subsection{Diverse and dynamic agentic workloads}
\label{sec:back-context}
\paragraph{Workload composition.}
An agent solves a task in steps orchestrated by a harness. Each LLM request
appends new tokens from tools or the harness to a prefix whose KV was computed in
earlier steps. The inference engine fetches the cached prefix KV, prefills the
uncached tokens, and decodes the output tokens. The amounts of prefill and decode work differ across tasks, harnesses, and
models. For example, the workload of a computer-use agent on OSWorld is
prefill-heavy because each step appends a new screenshot to the context, whereas
the workload of a math agent on Putnam is decode-heavy because each request
generates a long reasoning trace (\autoref{fig:intro-diversity}). The best PD split thus differs across agent workloads, and finding it requires costly offline profiling (\autoref{sec:eval-ablation}). The workload composition of a cluster also changes over time. User-triggered agent task starts follow workday rhythms, whereas automatically triggered task starts are more evenly distributed throughout the
day~\cite{zheng2026semanticsworkflowsinfrastructureunderstanding}. In production LLM serving, the mean input length of requests to one model rises by 13\% from midnight to afternoon, while the mean output length falls by 18\%~\cite{xiang2026servegen}. The best PD split thus shifts over time (\autoref{fig:intro-dynamicpd}). The PD split must therefore change over time to
avoid load imbalance between prefill and decode nodes.

\paragraph{Context management.} The harness typically extends the context of a branch with model
outputs and tool results, and consecutive requests therefore share prefixes.
Harnesses also manage context through forking, sub-agent spawning, and
compaction. \emph{Forking} creates new branches that reuse the current
context, while spawning a \emph{sub-agent} starts a separate branch with a
new context. \emph{Compaction} replaces the history of a branch with a
summary~\cite{openai2025codexmax,anthropic2026compaction}. Once compaction
completes, even recently accessed KV of the replaced history is no longer
needed. In contrast, KV can remain idle during a long wait for tools or
sub-agents, yet still be reused when the branch resumes. In GitHub Copilot,
the median gap between consecutive turns is 172~s~\cite{liu2026agentic}.
Whether KV will be reused thus depends on whether an active branch still
includes it in its context, not merely on access recency. Together
with the shift in workload composition, these dynamics challenge the planner, the
router, and the memory pool of existing serving systems, which the rest of
this section reviews.

\subsection{Existing PD disaggregated serving systems}
\label{sec:back-existsystem}

\paragraph{Engines.}
By default, vLLM and SGLang fix the phase of each engine at launch in PD
disaggregation mode~\cite{vllm2026nixl,sglang2026pd}. For example, each SGLang
engine builds only the scheduling loop and KV transfer path of its own phase. A
prefill engine queues requests and sends their KV to decode engines, whereas a
decode engine reserves KV slots, receives KV, and captures CUDA graphs for
decoding. Each engine also indexes cached KV in local state, such as the prefix
tree of a prefill engine. A phase switch thus requires relaunching the engine
with a new configuration and discards its cached KV.

\paragraph{Planners.}
A planner sets the PD split by switching nodes between prefill and decode as
compute demand changes. Phase switches in existing planners are costly. A
switch in Dynamo relaunches the engine and keeps the node out of service for
minutes~\cite{nvidia2026planner}. A switch in Arrow~\cite{wu2025arrow} and
xLLM~\cite{liu2025xllm} reassigns the node to the other phase without a relaunch
but strands its cached KV. The split also follows coarse signals, namely prefill
queue depth and decode KV utilization in Dynamo~\cite{nvidia2026planner}, and
TTFT and per-token latency targets in Arrow, xLLM, and
Llumnix~\cite{llumnix2026adaptivepd}. These planners predict TTFT from the full
input length~\cite{wu2025arrow,liu2025xllm,llumnix2026slo}, which counts cached
prefixes as prefill work and overestimates prefill demand.
Sub-agents spawned at once by the harness create a short burst of requests.
The burst raises queueing delay and TTFT but leaves the long-term balance
between prefill and decode workloads unchanged. These signals thus misestimate
demand and trigger frequent phase switches, each costing minutes of downtime
or the cached KV of the switched node (\autoref{sec:motivation}).


\paragraph{Routers.}
A router assigns each request to a prefill and decode node. For prefill,
the SGLang Model Gateway and the Dynamo router weigh prefix-cache hits against
node load~\cite{sglang2026sglang,nvidia2026routing}. The Dynamo router keeps a
cluster-wide prefix tree that records which KV blocks each node holds, and the
engines update the tree through KV events~\cite{nvidia2026routing}. For each
request, the router uses this tree to count the prefix blocks that each node
already holds. These routers choose where a request runs but do not move its KV.
Without a memory pool, a request thus either queues at a node that holds its
prefix, raising TTFT, or recomputes the prefix on another node, lowering
throughput.


\paragraph{Memory pools.}
A memory pool extends the capacity for reusable KV beyond GPU memory. HiCache
caches KV in GPU memory (L1) and host memory (L2) on each node and shares KV
across nodes through an optional storage tier (L3), such as Mooncake
Store~\cite{xie2025sglang,qin2025mooncake}. Shared KV moves from the L2 of the
source node through L3 to the L2 of the destination node, leaving copies in L3
and in the L2 of both nodes. HiCache loads the fetched KV into GPU memory only
when the request starts its prefill~\cite{sglang2026hicache}. This load
therefore delays prefill and raises TTFT (\autoref{sec:motiv-kv}). These pools
index KV by content and typically evict KV in LRU order. LRU eviction thus cannot
distinguish the KV of a branch that waits for a tool from the history that
compaction has replaced. Agent-aware systems such as Continuum and ThunderAgent
retain the KV of agents that wait for tools~\cite{li2025continuum,kang2026thunderagent},
but assume that each context only grows and do not track the duplicate copies
that KV transfers leave across nodes.

In summary, these systems couple compute allocation with KV cache management,
making phase switches costly. Their planners rely on coarse signals that do not
fully reveal the workload composition, and their routers trade cache locality
against load balance. Their memory pools load fetched KV only when prefill
starts, evict KV by access recency, and leave duplicate copies after KV
transfers. Agent-aware systems retain the KV of waiting agents but assume that
contexts only grow. \autoref{sec:motivation} quantifies the resulting losses.

\section{Motivation}
\label{sec:motivation}

This section profiles PD-disaggregated serving systems on agentic
workloads and quantifies three sources of inefficiency, namely costly phase
switches, misestimated compute demand, and misjudged KV reuse. We use the
testbed of \autoref{sec:eval-setup}, nine nodes of 4$\times$GB300 GPUs that
serve GLM-5.2 with TP4, and replay the Claude Code traces with
90 concurrent agents for one hour after a 30-minute
warmup. Dynamo runs its planner at the default throughput target with the
SGLang Model Gateway. Arrow's code is not public, so we reimplement its
planner and scheduler on the symmetric engines of \sys{}
(\autoref{sec:dataplane-switch}). Both planners start from 6P3D, and each node
runs HiCache with LRU eviction. For routing (\autoref{sec:motiv-kv}), we run SGLang in its PD disaggregation mode at the best static split, 4P5D, and compare the cache-aware and sticky-session policies of the SGLang Model Gateway, the Dynamo KV router~\cite{nvidia2026routing}, and a variant of the Dynamo KV router whose nodes fetch prefixes from a shared host-memory pool (+ memory pool).

\begin{figure}[t] %
  \centering
  \begin{subfigure}[t]{0.325\textwidth}
    \centering
    \includegraphics[width=\linewidth]{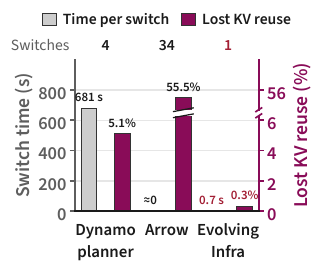}
    \vspace{-7mm}
    \caption{Costly phase switches}
    \label{fig:motiv-cost}
  \end{subfigure}\hfill
  \begin{subfigure}[t]{0.325\textwidth}
    \centering
    \includegraphics[width=\linewidth]{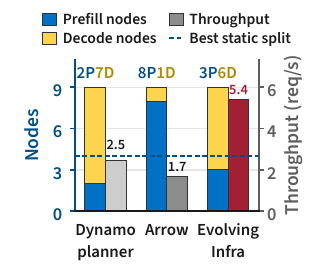}
    \vspace{-7mm}
    \caption{Misestimated compute demand}
    \label{fig:motiv-compute}
  \end{subfigure}\hfill
  \begin{subfigure}[t]{0.325\textwidth}
    \centering
    \includegraphics[width=\linewidth]{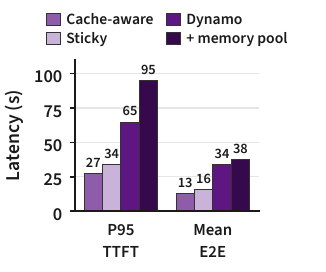}
    \vspace{-7mm}
    \caption{Misjudged KV reuse}
    \label{fig:motiv-memory}
  \end{subfigure}
  \vspace{-1mm} %
  \caption{\textbf{Costly phase switches, misestimated compute demand and misjudged KV reuse
  in PD disaggregation systems.} Lost KV reuse is $1 - h/h^*$, with $h$ the KV hit
  rate and $h^*$ that with an unlimited KV cache shared by all nodes and no phase switches. (a) Switch time and lost KV reuse; once its running requests finish, a \sys{} node switches in 0.7\,s, and \sys{} loses 0.3\% of KV reuse. (b) Final split and throughput; \sys{}
  achieves over twice the throughput of either planner.
  (c) P95 TTFT and mean E2E latency at the best static split, 4P5D; + memory pool adds shared host memory to Dynamo.}
  \label{fig:motiv}
\end{figure}

\subsection{Costly phase switches}
\label{sec:motiv-switch}
Every phase switch takes time, loses KV reuse, or both
(\autoref{fig:motiv-cost}), because engines fix their phase at launch and
index cached KV locally (\autoref{sec:back-existsystem}). The Dynamo planner relaunches the worker of a
node at every switch. A switch takes \textbf{681}~s on average, during which the node serves no requests. Arrow switches nodes in place without this delay, but the KV
of a switched node becomes unreachable for prefill. Arrow misses 55.5\% of the cache hits that it would achieve with an unlimited KV cache shared by all nodes and no phase switches, and Dynamo misses 5.1\%. We call this fraction of missed hits the
\emph{lost KV reuse}.
Avoiding both costs requires an engine that serves both phases without a
relaunch and a KV pool that every node can read in either phase.

\subsection{Misestimated compute demand}
\label{sec:motiv-demand}
Both online planners size the split from coarse signals
(\autoref{sec:background}). The Dynamo planner and Arrow thus deliver 49\%
and 65\% less throughput than the best static split (4.9~req/s), respectively
(\autoref{fig:motiv-compute}). These signals misjudge the work of agents in
opposite directions. Arrow predicts prefill time from the full input length,
which counts cached prefixes as work, and it also reacts to short prefill
bursts, such as several sub-agents that start together. Arrow thus drifts
from 6P3D to 8P1D, and switches 34 times during the run, each time incurring the costs in \autoref{sec:motiv-switch}. The Dynamo
planner instead follows queue depth and KV utilization. Long agent contexts
keep the KV utilization of decode nodes high, and the planner drifts past
4P5D to 2P7D. Sizing the split well requires a model of the average prefill and decode work
that a request needs after KV reuse. Short bursts of sub-agents and forks raise the momentary load but barely change the average.

\subsection{Misjudged KV reuse}
\label{sec:motiv-kv}

Routers misjudge the KV reuse of agents in either direction
(\autoref{fig:motiv-memory}). The cache-aware and
sticky policies of the SGLang Model Gateway keep an agent and its forked
branches on the few prefill nodes that hold their prefixes. The P95 TTFT
thus reaches 27~s and 34~s. The load-aware Dynamo KV
router~\cite{nvidia2026routing} sends more requests to nodes that lack their
prefixes, which recompute them, and the P95 TTFT reaches 65~s. With the
shared memory pool, these nodes fetch the prefixes instead, yet the P95 TTFT
rises to 95~s. Because the pool loads the fetched KV into GPU memory only when the request
begins its prefill, the load adds to the prefill time instead of overlapping
with queueing (\appref{app:exp-async-fetch}). The duplicate copies that fetches leave also
occupy host memory and displace the KV that agents still need.

Eviction misjudges KV reuse as well. HiCache evicts the least recently used
KV, whereas the reuse of agent KV depends on whether an active branch still
includes it (\autoref{sec:background}). HiCache thus keeps histories that
compaction replaced but evicts the KV of branches that wait for tools,
sub-agents, or users. On the CC traces, these evictions cause 39\% of the
KV reuse that Dynamo loses (\appref{app:exp-switch}). Reusing KV well requires
a memory pool that hides fetch delays, routing that balances load, and
eviction that follows agent context management and removes duplicate copies.

\section{Decoupled and Agent-Aware Infrastructure}
\label{sec:dataplane}

This section presents the infrastructure that enables efficient online
adaptation (\autoref{fig:teaser}). The symmetric engine
(\autoref{sec:dataplane-switch}) allows nodes to switch phases without a
relaunch, and the shared KV pool (\autoref{sec:dataplane-pool}) keeps KV
readable by all nodes in either phase. Together, they meet the two needs in
\autoref{sec:motiv-switch} and decouple compute allocation and KV cache
management. The KV directory (\autoref{sec:dataplane-directory}) counts the
active branches that reference each KV block on each node and updates these
counts as branches fork, compact, or finish and as KV moves between nodes. These counts
identify the KV that no active branch reuses (\autoref{sec:motiv-kv}).

\subsection{Symmetric engines for online phase switches}
\label{sec:dataplane-switch}

To reduce the cost of online prefill--decode phase switches
(\autoref{sec:motiv-switch}), we build a symmetric engine that supports both
phases, so that every node can serve either phase. With
this engine, the router separates prefill and decode at runtime rather than
through a static deployment configuration. Unlike SGLang (\autoref{sec:background}), whose prefill
and decode engines run different scheduling loops and fix their phase at launch
by default, our engine extends SGLang with one scheduling loop that serves both
phases and a KV transfer protocol that moves KV between any two nodes in either
direction. The protocol reuses the RDMA transport of
Mooncake~\cite{qin2025mooncake} and keeps working after a node switches its
phase. Every node of \sys{} runs this engine, and a single router routes prefill
requests to one set of nodes and decode requests to the remaining nodes
(\autoref{fig:teaser}). When the planner reassigns the phase of a node
(\autoref{sec:design-split}), the router changes the type of requests that it
routes to the node, and the engine serves the new phase without a restart.


\subsection{Shared host-memory KV pool}
\label{sec:dataplane-pool}

To keep KV readable by every node in either phase (\autoref{sec:motiv-switch}),
\sys{} pools the host memory (L2) of all nodes into a shared KV pool. Following
HiCache~\cite{xie2025sglang}, each node caches KV in GPU memory (L1) and writes
it through to its L2 in both phases. A prefill node locates reusable KV through
the directory (\autoref{sec:dataplane-directory}) and pulls it from the L2 of a
peer over RDMA. Unlike the storage tier of SGLang, which stages shared KV in a
separate store, the pool reads directly from the L2 of a peer and avoids this
extra copy (\appref{app:related-kvstruct}). Deduplication across the GPUs of a
node further increases its effective capacity (\appref{app:cpu-dedup}). Decode
nodes only write to the pool, since a fetch from host memory would delay
decoding (\appref{app:exp-async-fetch}). They receive the KV of each request
from the prefill node through the faster GPU-to-GPU transfer, and the KV that
they generate stays in their L2 for the next request of the branch.

A fetch from a peer, however, delays the prefill of its request
(\autoref{sec:motiv-kv}). HiCache prefetches KV only into host memory while a request waits and loads it
into GPU memory layer by layer during prefill. \sys{} instead completes the whole
fetch while the request waits in the queue. The node pulls the KV from the L2 of
the peer into its own L2, loads it into GPU memory, and broadcasts it to all GPUs
of the node before the prefill begins.

When the planner selects a new split (\autoref{sec:design-split}), the router
moves the queued requests of each switching node to other nodes of its current
phase. The running requests of the node finish in place, after which the node
switches its phase. The KV of the switching node stays in the pool and remains
readable by all nodes throughout the switch (\autoref{sec:motiv-switch}).

\subsection{Agent-aware cluster-wide KV directory}
\label{sec:dataplane-directory}

\begin{figure*}[t]
  \centering
  \captionsetup[sub]{justification=centering}
  \input{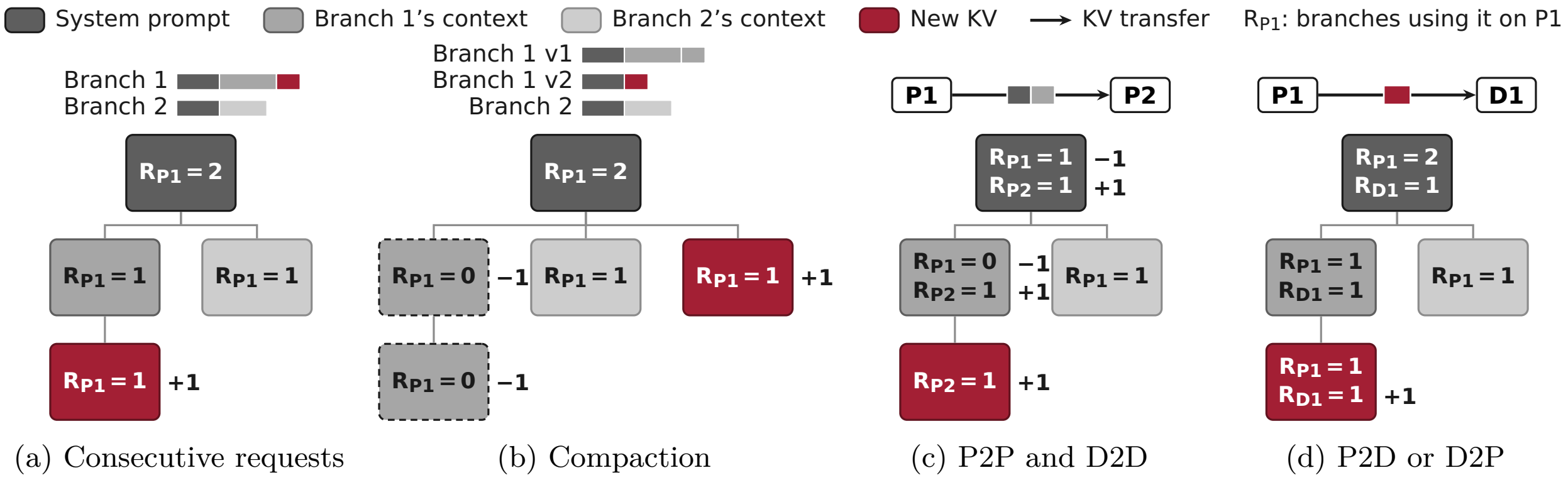}%
  \resizebox{\linewidth}{!}{\begin{tikzpicture}[dir, sw/.style={minimum width=0.42cm, minimum height=0.26cm, anchor=west}]
  \node[dirsys, sw] (k1) at (0,0) {};
  \node[dirlab, anchor=west] (t1) at ([xshift=3pt]k1.east) {System prompt};
  \node[dirh1, sw] (k2) at ([xshift=0.35cm]t1.east) {};
  \node[dirlab, anchor=west] (t2) at ([xshift=3pt]k2.east) {Branch 1's context};
  \node[dirh2, sw] (k3) at ([xshift=0.35cm]t2.east) {};
  \node[dirlab, anchor=west] (t3) at ([xshift=3pt]k3.east) {Branch 2's context};
  \node[dirnew, sw] (k4) at ([xshift=0.35cm]t3.east) {};
  \node[dirlab, anchor=west] (t4) at ([xshift=3pt]k4.east) {New KV};
  \draw[dirmove, ->] ([xshift=0.35cm]t4.east) -- ++(0.45,0) coordinate (k6);
  \node[dirlab, anchor=west] (t6) at ([xshift=3pt]k6) {KV transfer};
  \node[dirlab, anchor=west] at ([xshift=0.35cm]t6.east)
    {R\kern0.02em\raisebox{-0.3ex}{\fontsize{5.4}{6}\selectfont P1}: branches using it on P1};
\end{tikzpicture}}\par\vspace{2pt}%
  \begin{subfigure}[t]{0.225\textwidth}%
    \centering
    \begin{tikzpicture}[dir]
  \dirframe
  \dirbar{1.555}{3.66}{Branch 1}{eiGrayD/0.45, eiGrayM/0.6, eiRed/0.25}
  \dirbar{1.555}{3.38}{Branch 2}{eiGrayD/0.45, eiGrayL/0.5}
  \dirblk{s}{1.6}{\dirtop}{dirsys}{\dirR{P1}{2}}
  \dirblk{h1}{0.8}{\dirmid}{dirh1}{\dirR{P1}{1}}
  \dirblk{h2}{2.4}{\dirmid}{dirh2}{\dirR{P1}{1}}
  \dirblk{n}{0.8}{\dirbot}{dirnew}{\dirR{P1}{1}}
  \dirdown{s}{h1} \dirdown{s}{h2} \draw[diredge] (h1) -- (n);
  \dirup{n}{0}
\end{tikzpicture}%
    \caption{Consecutive requests}
    \label{fig:directory-consecutive}
  \end{subfigure}\hfill
  \begin{subfigure}[t]{0.30\textwidth}%
    \centering
    \begin{tikzpicture}[dir]
  \dirframe
  \dirbar{1.555}{3.94}{Branch 1 v1}{eiGrayD/0.45, eiGrayM/0.6, eiGrayM/0.25}
  \dirbar{1.555}{3.66}{Branch 1 v2}{eiGrayD/0.45, eiRed/0.25}
  \dirbar{1.555}{3.38}{Branch 2}{eiGrayD/0.45, eiGrayL/0.5}
  \dirblk{s}{1.6}{\dirtop}{dirsys}{\dirR{P1}{2}}
  \dirblk{h1}{0}{\dirmid}{dirh1, dirzero}{\dirR{P1}{0}}
  \dirblk{h2}{1.6}{\dirmid}{dirh2}{\dirR{P1}{1}}
  \dirblk{m}{3.2}{\dirmid}{dirnew}{\dirR{P1}{1}}
  \dirblk{n}{0}{\dirbot}{dirh1, dirzero}{\dirR{P1}{0}}
  \dirdown{s}{h1} \dirdown{s}{h2} \dirdown{s}{m} \draw[diredge] (h1) -- (n);
  \dirminus{h1}{0} \dirminus{n}{0} \dirup{m}{0}
\end{tikzpicture}%
    \caption{Compaction}
    \label{fig:directory-compaction}
  \end{subfigure}\hfill
  \begin{subfigure}[t]{0.225\textwidth}%
    \centering
    \begin{tikzpicture}[dir]
  \dirframe
  \dirxfer{P1}{P2}{->}{1.35}{eiGrayD/0.25, eiGrayM/0.25}
  \dirblkii{s}{1.6}{\dirtop}{dirsys}{\dirR{P1}{1}}{\dirR{P2}{1}}
  \dirblkii{h1}{0.8}{\dirmid}{dirh1}{\dirR{P1}{0}}{\dirR{P2}{1}}
  \dirblk{h2}{2.4}{\dirmid}{dirh2}{\dirR{P1}{1}}
  \dirblk{n}{0.8}{\dirbot}{dirnew}{\dirR{P2}{1}}
  \dirdown{s}{h1} \dirdown{s}{h2} \draw[diredge] (h1) -- (n);
  \dirminus{s}{1} \dirup{s}{2} \dirminus{h1}{1} \dirup{h1}{2} \dirup{n}{0}
\end{tikzpicture}%
    \caption{P2P and D2D}
    \label{fig:directory-p2p}
  \end{subfigure}\hfill
  \begin{subfigure}[t]{0.225\textwidth}%
    \centering
    \begin{tikzpicture}[dir]
  \dirframe
  \dirxfer{P1}{D1}{->}{1.46}{eiRed/0.28}
  \dirblkii{s}{1.6}{\dirtop}{dirsys}{\dirR{P1}{2}}{\dirR{D1}{1}}
  \dirblkii{h1}{0.8}{\dirmid}{dirh1}{\dirR{P1}{1}}{\dirR{D1}{1}}
  \dirblk{h2}{2.4}{\dirmid}{dirh2}{\dirR{P1}{1}}
  \dirblkii{n}{0.8}{\dirbot}{dirnew}{\dirR{P1}{1}}{\dirR{D1}{1}}
  \dirdown{s}{h1} \dirdown{s}{h2} \draw[diredge] (h1) -- (n);
  \dirup{n}{2}
\end{tikzpicture}%
    \caption{P2D or D2P}
    \label{fig:directory-p2d}
  \end{subfigure}
  \caption{\textbf{Counts follow agent context management and KV transfers.} Each block of the
  cluster-wide prefix tree shows, for each node that holds it, the number of active branches
  that reference it there: R\textsubscript{P1} and R\textsubscript{P2} on prefill nodes P1 and P2,
  and R\textsubscript{D1} on decode node D1. Tags $+1$ and $-1$ mark the counts that the step
  changes, and both branches start on P1.
  (a)~A consecutive request counts only its new block, and reused blocks keep their counts.
  (b)~Compaction increments the version of branch~1 from v1 to v2, so its old blocks drop to
  R\textsubscript{P1}\,=\,0 (dashed). (c)~When branch~1 moves from P1 to P2, P2 counts its KV and P1
  releases its counts. (d)~When branch~1 alternates between P1 and D1, the target nodes count its KV,
  and each step sends only new KV.}
  \label{fig:directory}
  \vspace{-3mm}
\end{figure*}

To identify the duplicate copies that KV transfers leave and the histories
that context management drops (\autoref{sec:motiv-kv}), the router
(\autoref{fig:teaser}) maintains a cluster-wide KV directory. Following the
Dynamo KV indexer~\cite{nvidia2026routing}, the directory is a prefix tree
that records KV placement in GPU and host memory of each node. The router
updates this tree with the KV events that the engines publish
(\autoref{sec:background}). The directory also records, for each block on each node, how many active branches reference it (\autoref{fig:directory}).

The router updates these counts from two identifiers. The harness tags each
request with the ID of the agent branch that issues it and reports when a
branch finishes. The router maintains a version for each branch and
increments it when the context of the branch is rewritten. The engine
on each node then updates the eviction priority of the KV blocks in its host
memory according to the counts that the router sends
(\autoref{sec:design-placement}).

Both identifiers follow agent context management. When the harness forks a
context or spawns a sub-agent, it sends the ID of the new branch to the
router. The router then adds one count to each block that the new branch
reuses on the node that serves it. For consecutive requests of the same
branch, the router compares the new request with the previous request of the
branch. If the new request only appends tokens, the router adds counts only
to the newly appended blocks on the routed node
(\autoref{fig:directory-consecutive}). The reused blocks keep their counts,
since the branch was already counted on them in earlier steps. If the
context is rewritten by compaction, the router increments the version. It
then releases the counts on the blocks that the new context drops and adds
counts to the blocks that it newly includes
(\autoref{fig:directory-compaction}). When the harness reports that a branch
has finished, the router releases all counts of the branch.

Counts also follow KV transfers. A transfer between two nodes of the same
phase (P2P or D2D) moves the counts of the branch, since the branch reuses
only the target copy in that phase. The source releases them once the target
holds the current version (\autoref{fig:directory-p2p}). A transfer across
phases (P2D or D2P) keeps the counts on both nodes, since the next step of
the branch needs the copies in both phases (\autoref{fig:directory-p2d}). A
node that switches from decode to prefill resets its counts to free its
memory, whereas a node that switches to decode keeps them to preserve its KV.
\appref{app:details-branch} and \appref{app:details-refcount}  give these
rules in pseudocode. As a result, each active branch holds counts on at most one copy per phase.

\vspace{-4mm}
\section{Agent-Adaptive Compute Allocation and KV Cache Management}
\label{sec:design}
Based on the infrastructure in \autoref{sec:dataplane}, this section presents
how \sys{} adapts compute allocation and KV cache management to agents at
runtime. We first characterize the work of each request and learn
iteration-time models of each node online (\autoref{sec:design-model}).
\autoref{sec:design-split} then presents the planner, which uses these models
to size the PD split for diverse agent workloads and their shifting
composition. Given the split, \autoref{sec:design-placement} presents
KV-reuse-aware eviction and routing, which preserve the KV that active branches still reuse and balance prefill work across nodes. \autoref{tab:notation}
summarizes the notation.

\begin{table}[htb]%
\centering
\small
\caption{Summary of notation for requests and the cluster.}
\label{tab:notation}
\begin{tabular}{@{}ll@{}}
\toprule
Notation & Description \\
\midrule
$t_{\mathrm{uc}}$, $t_{\mathrm{c}}$ & Uncached and cached prefill tokens of a request \\
$t_{\mathrm{out}}$, $t_{\mathrm{acc}}$ & Output tokens; mean accepted tokens per decode iteration \\
$t_{\mathrm{fwd}}$ & Tokens that each running request feeds into a decode iteration \\
$w$ & Work of a request, defined by $(t_{\mathrm{uc}}, t_{\mathrm{c}}, t_{\mathrm{out}}, t_{\mathrm{acc}})$ \\
$N$, $n$ & Number of nodes; index of a node \\
$\mathcal{P}$, $\mathcal{D}$ & Sets of prefill and decode nodes \\
$B$ & Number of running requests on a decode node \\
\bottomrule
\end{tabular}
\end{table}

\begin{figure*}[t]
  \centering
  \captionsetup[sub]{justification=centering}
  \begin{subfigure}[t]{0.232\textwidth}%
    \centering
    \includegraphics[width=\linewidth]{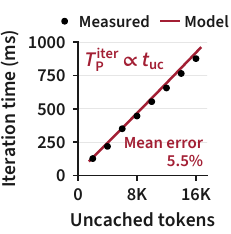}
    \captionsetup{margin={0.03in,0in}}  %
    \caption{Prefill iteration time\\ vs.\ uncached tokens}
    \label{fig:itmodel-uncached}
  \end{subfigure}\hfill
  \begin{subfigure}[t]{0.232\textwidth}%
    \centering
    \includegraphics[width=\linewidth]{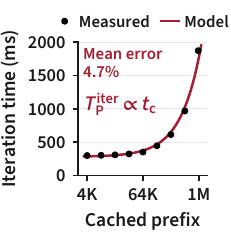}
    \captionsetup{margin={0.03in,0in}}%
    \caption{Prefill iteration time\\ vs.\ cached prefix}
    \label{fig:itmodel-prefix}
  \end{subfigure}\hfill
  \begin{subfigure}[t]{0.232\textwidth}%
    \centering
    \includegraphics[width=\linewidth]{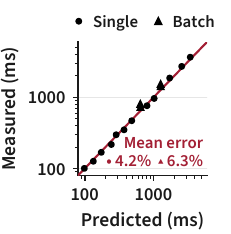}
    \captionsetup{margin={0.03in,0in}}%
    \caption{Predicted vs.\\ measured iteration time}
    \label{fig:itmodel-all}
  \end{subfigure}\hfill
  \begin{subfigure}[t]{0.30\textwidth}%
    \centering
    \includegraphics[width=\linewidth]{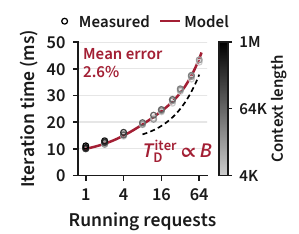}
    \captionsetup{margin={0in,0.05in}}
    \caption{Decode iteration time\\ vs.\ running requests}
    \label{fig:itmodel-decode}
  \end{subfigure}
  \caption{\textbf{The iteration-time models closely track the measured iteration
times.} Points are GPU times per iteration of GLM-5.2 on four GB300 GPUs, and red
lines are the models in Equations~\eqref{eq:prefill-step}
and~\eqref{eq:decode-model}. We fit the prefill model on 20 single-request
settings and the decode model on the ten points with a 16K context, and each
panel reports the mean error on the remaining points. (a)~The prefill iteration
time grows nearly linearly with the uncached prefill tokens at a 128K cached
prefix. (b)~It grows linearly with the cached prefix at 8K uncached prefill
tokens (log-scale axis). (c)~Outside the fitted settings, the mean error is
4.2\% for single requests and 6.3\% for batches of 2--8 requests. The panel shows
16 of the 60 measured prefill iterations, evenly spaced and including the four
batches with the largest errors. (d)~The decode iteration time follows the model
as the running requests grow, and with 8 or more running requests, the context
length (darker for longer) changes it by only 2--7\%.}
  \label{fig:itmodel}
  \vspace{-4mm}
\end{figure*}

\subsection{Prefill and decode iteration-time models}
\label{sec:design-model}

As \autoref{sec:motiv-demand} shows, sizing the split requires a model of the
average prefill and decode work that a request needs after KV reuse. This
section models how long a request with given work occupies a node, and
\autoref{sec:design-split} evaluates this model on the average work over each
planning window. A node runs
its requests in iterations, each of which is a forward pass of the model.
Under continuous batching~\cite{yu2022orca}, each iteration serves a batch of
requests, and requests join or leave the batch between iterations. Accurate
predictions of iteration time thus allow \sys{} to estimate the prefill and
decode throughput of each node and the waiting time of prefill requests. For
each phase, we model the iteration time from the requests in its batch and derive the number of requests that each node completes per
second. We describe each request in a batch by its work
$w = (t_{\mathrm{uc}}, t_{\mathrm{c}}, t_{\mathrm{out}}, t_{\mathrm{acc}})$,
where \autoref{tab:notation} defines each component. The router obtains
$t_{\mathrm{c}}$ from the longest cached prefix that the KV directory finds
for the request (\autoref{sec:dataplane-directory}), and $t_{\mathrm{uc}}$
counts the remaining input tokens. Separating the two token lengths allows
the models to predict prefill time accurately, even when agents reuse most
of their context.

The models rest on two premises. First, the kernels of a forward pass run
one after another, and the iteration time is the sum of their execution
times. Second, as in the roofline model~\cite{williams2009roofline}, each
kernel is limited either by the compute throughput or by the memory
bandwidth of the GPU. A compute-bound kernel runs in time proportional to
its arithmetic operations, whereas a memory-bound kernel runs in time
proportional to the bytes that it reads and writes.

\textbf{Prefill iteration-time model.} Prefill spends most of its time in
compute-bound kernels~\cite{zhong2024distserve}. Under chunked
prefill~\cite{agrawal2024tamingthroughputlatencytradeoffllm}, each prefill
iteration processes the uncached tokens of several requests up to a fixed
token budget. The execution time of the linear layers grows in proportion to
the uncached tokens of the batch, and that of the attention grows in proportion
to its query--key pairs. Under causal attention, the $i$-th uncached token attends to the $t_{\mathrm{c}}$
cached tokens and to the first $i$ uncached tokens, including itself. A request
therefore computes
$A = t_{\mathrm{uc}} t_{\mathrm{c}} + t_{\mathrm{uc}}(t_{\mathrm{uc}}+1)/2$
query--key pairs (\autoref{fig:itmodel-prefix}). A request reuses its cached prefix from GPU memory or from host memory.
Asynchronous fetches hide most of the restore from host memory
(\autoref{sec:dataplane-pool}), but each restored token still adds a small cost to the iteration, such as the remaining host-to-GPU copy. This cost grows in proportion to the restored tokens. Each
iteration also pays a fixed cost, such as the time to launch its kernels. A
prefill iteration on node $n$ thus takes
\begin{equation}
  T^{\mathrm{iter}}_{\mathrm{P},n} = c_{n,0}
    + c_{n,1} \textstyle\sum t_{\mathrm{uc}}
    + c_{n,2} \textstyle\sum A
    + c_{n,3} \textstyle\sum t_{\mathrm{rst}},
  \label{eq:prefill-step}
\end{equation}
where the sums run over the requests of the batch, and $t_{\mathrm{rst}}$ is
the number of tokens that a request restores from host memory. The coefficients $c_{n,0}$ to $c_{n,3}$ give the fixed cost of an iteration and
the time per uncached token, per query--key pair, and per restored token. They depend on the model, hardware, and parallelism of node $n$,
and \sys{} fits them \textbf{online}, as described at the end of this section.
\appref{app:theory-attention} shows that Equation~\eqref{eq:prefill-step} also
holds for sliding-window, linear, and sparse attention, including the DeepSeek
Sparse Attention (DSA) of GLM-5.2.

\textbf{Decode iteration-time model.} Decode spends most of its time in
memory-bound kernels~\cite{zhong2024distserve}, whose execution time grows
in proportion to the bytes they read. Most of these bytes belong to the
model weights. Each iteration reads the dense weights in full, which forms a
fixed cost. In the mixture-of-experts (MoE) layers, the iteration reads only
the weights of the routed experts that its tokens select. A decode iteration
serves the $B$ running requests of a node, and each request feeds
$t_{\mathrm{fwd}}$ tokens into the iteration, a number that exceeds one
under speculative decoding. With each token selecting $k$ of the $E$ routed
experts, the $t_{\mathrm{fwd}} B$ tokens select an expected fraction
$\phi(B) = 1 - (1 - k/E)^{t_{\mathrm{fwd}} B}$ of them
(\appref{app:theory-moe}). A decode iteration holds only a few tokens, and this fraction therefore changes with $B$. Each running
request also adds about the same time to the iteration, for example to
sample its next tokens and to draft tokens for speculative decoding. This
part of the iteration time grows in proportion to $B$. The time to read the
KV of running requests also grows in proportion to their total context
length. GLM-5.2 and Kimi K2.6, however, store little KV per token and read only part of
it under sparse attention (\appref{app:theory-context}). At 8 or more
running requests, the context length thus changes the iteration time by
only 2--7\% (\autoref{fig:itmodel-decode}), and the model omits this term. A
decode iteration on node $n$ thus takes
\begin{equation}
  T^{\mathrm{iter}}_{\mathrm{D},n}(B) = d_{n,0} + d_{n,1}\,\phi(B) + d_{n,2}\, B.
  \label{eq:decode-model}
\end{equation}
The coefficients $d_{n,0}$ to $d_{n,2}$ give the fixed cost of an iteration,
the time to read the weights of all routed experts, and the time per running
request. Like the prefill coefficients, they depend on the model, hardware,
and parallelism of node $n$, and \sys{} fits them \textbf{online}.

\textbf{Request time and node throughput.} In
Equation~\eqref{eq:prefill-step}, every term except the fixed cost sums over
the requests of the batch. A request thus occupies the prefill node for its
own share of these terms plus part of the fixed cost, which we divide among
the requests in proportion to their uncached tokens. With $\bar{t}_n$ the average
number of new tokens per prefill iteration on node $n$ and $f_n$ the
fraction of cached tokens that the node restores from host memory, a request
with work $w$ occupies node $n$ for
\begin{equation}
  T^{\mathrm{req}}_{\mathrm{P},n}(w) = c_{n,0}\, t_{\mathrm{uc}}/\bar{t}_n
    + c_{n,1}\, t_{\mathrm{uc}} + c_{n,2}\, A + c_{n,3}\, f_n\, t_{\mathrm{c}}
  \label{eq:prefill-model}
\end{equation}
in prefill. In decode, each iteration advances every running request by
$t_{\mathrm{acc}}$ tokens on average, and a request needs
$t_{\mathrm{out}}/t_{\mathrm{acc}}$ iterations to finish. Since the $B$
running requests share each iteration, a request occupies node $n$ for
\begin{equation}
  T^{\mathrm{req}}_{\mathrm{D},n}(w) = T^{\mathrm{iter}}_{\mathrm{D},n}(B)\,
    t_{\mathrm{out}} \big/ \bigl(B\, t_{\mathrm{acc}}\bigr)
  \label{eq:decode-req}
\end{equation}
in decode. Node $n$ thus completes
$\mu_{\mathrm{P},n} = 1/T^{\mathrm{req}}_{\mathrm{P},n}(w)$ requests per
second in prefill and $\mu_{\mathrm{D},n} = 1/T^{\mathrm{req}}_{\mathrm{D},n}(w)$
requests per second in decode.

\textbf{Online fitting.} \sys{} fits both models purely online at the start of
serving. During a warmup period, every node serves both phases in a collocated
mode, and \sys{} records the prefill and decode work that each node completes in fixed windows. It fits the coefficients of both models for each node by
nonnegative least squares, freezes them, and then starts PD disaggregation from the initial split. After warmup, \sys{} periodically remeasures only $\bar{t}_n$ and $f_n$, which change with the workload composition
(\autoref{fig:intro-dynamicpd}). On controlled measurements on GB300, the model form predicts iteration time with a mean error below 7\% outside the fitted settings (\autoref{fig:itmodel}).

\vspace{-3mm}
\subsection{Online phase-split adaptation}
\label{sec:design-split}

To size the PD split for diverse agent workloads and their shifting composition
(\autoref{fig:intro-diversity} and \autoref{fig:intro-dynamicpd}), the planner
reevaluates the possible splits by their predicted throughput at the end of each
planning window of length $T_{\mathrm{win}}$. Each request passes through both phases, and the slower phase
thus determines the cluster throughput,
\begin{equation}
  \Theta(\mathcal{P}, \mathcal{D}; \bar{w}, \bar{q}) =
  \min\Bigl( \sum_{n\in\mathcal{P}} \mu_{\mathrm{P},n},\;
             \sum_{n\in\mathcal{D}} \mu_{\mathrm{D},n} \Bigr),
  \label{eq:throughput}
\end{equation}
where $\mathcal{P}$ and $\mathcal{D}$ together contain the $N$ nodes of the
cluster. Here $\bar{w}$ is the average work of a request over the window,
and $\bar{q}$ is the average number of requests in decode. Unlike queue depth, $\bar{w}$ characterizes the agent workload itself and barely
changes during short bursts of sub-agents and forks. The planner evaluates
$\mu_{\mathrm{P},n}$ with $\bar{w}$ in Equation~\eqref{eq:prefill-model} and
$\mu_{\mathrm{D},n}$ with the batch $B = \bar{q}/|\mathcal{D}|$ in
$T^{\mathrm{req}}_{\mathrm{D},n}$.

Since each switch has a cost (\autoref{sec:motiv-switch}), the planner
switches only when another split is statistically better than the current
one. For each prefill count other than the current one, the planner forms a
candidate split $(\mathcal{P}', \mathcal{D}')$. It divides the window into
$m$ slices and measures $\bar{w}_i$ and $\bar{q}_i$ on each slice $i$. The
gain of the candidate on slice $i$ is
\begin{equation}
  g_i = \Theta(\mathcal{P}', \mathcal{D}'; \bar{w}_i, \bar{q}_i)
      - \Theta(\mathcal{P}, \mathcal{D}; \bar{w}_i, \bar{q}_i).
  \label{eq:gain}
\end{equation}
The candidate qualifies if its mean gain exceeds twice its standard error,
$\bar{g} > 2\sigma_g/\sqrt{m}$, where $\bar{g}$ and $\sigma_g$ are the mean
and standard deviation of the gains $g_i$. This threshold keeps short
bursts from triggering switches (\autoref{sec:motiv-demand}). If any
candidate qualifies, the planner switches to the one with the largest
$\bar{g}$. It applies the new split through the phase switch in
\autoref{sec:dataplane-switch}, and \appref{app:details-planner} summarizes the
planner.

\vspace{-2mm}
\subsection{KV-reuse-aware eviction and routing}
\label{sec:design-placement}

\textbf{Eviction.} The rules in \autoref{sec:dataplane-directory} keep the
count of a copy positive only while an active branch may still reuse it on
that node. A block with a zero count, such as the history that compaction replaced, the context of a finished sub-agent, or the source copy left by a same-phase transfer (\autoref{fig:directory}), is thus KV that no active branch will reuse.
Each node therefore evicts the blocks in its host memory in ascending order
of the key $(\mathds{1}[R_b > 0],\, T_b)$, where $R_b$ is the count of
block $b$ on the node and $T_b$ is its last access time. The node thus
evicts zero-count blocks first, in LRU order, and evicts blocks with
positive counts only after no zero-count block remains. Unlike LRU (\autoref{sec:motiv-kv}), this order evicts replaced histories before
the KV of branches that wait for tools, sub-agents, or users.

\textbf{Routing.} With asynchronous fetches (\autoref{sec:dataplane-pool}) and
the eviction policy above, KV transfers are fast and memory-efficient. The router
therefore balances the predicted TTFT across prefill nodes. For a request with
work $w$ and prefill node $n$, the router predicts
\begin{equation}
  \hat{T}_n(w) = T^{\mathrm{rem}}_n
    + \textstyle\sum_{j \in Q_n} T^{\mathrm{req}}_{\mathrm{P},n}(w_j)
    + T^{\mathrm{req}}_{\mathrm{P},n}(w),
  \label{eq:ttft}
\end{equation}
where $T^{\mathrm{rem}}_n$ is the remaining time of the running batch on node $n$,
$Q_n$ is the set of its queued requests, and each service time follows
Equation~\eqref{eq:prefill-model}. \appref{app:details-planner} details how the
prediction also charges the part of a fetch that outlasts the wait and the
duplicate copy that the fetch creates. The router routes the request to the node
with the smallest $\hat{T}_n(w)$, and that node fetches the KV of the request
from the shared pool while the request waits. Unlike cache-aware routing
(\autoref{sec:motiv-kv}), this choice weighs the queue of each node against the
KV that the node must fetch and avoids overloading nodes with popular prefixes.

\begin{figure}[t]
  \centering
  \captionsetup[subfigure]{skip=2pt}
  \includegraphics{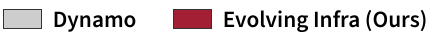}\\[1pt]
  \begin{subfigure}[t]{2.36in}
    \centering
    \captionsetup{margin={0.66in,0.10in}}
    \includegraphics{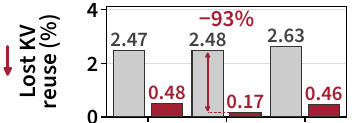}\\
    \includegraphics{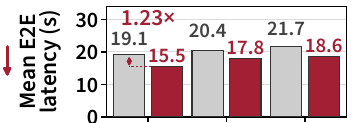}\\
    \includegraphics{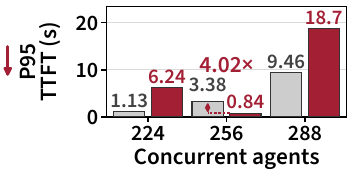}
    \caption{Putnam (GLM-5.2)}
    \label{fig:eval-single-putnam}
  \end{subfigure}\hspace{0.04in}%
  \begin{subfigure}[t]{1.94in}
    \centering
    \captionsetup{margin={0.29in,0.05in}}
    \includegraphics{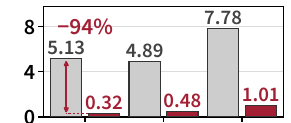}\\
    \includegraphics{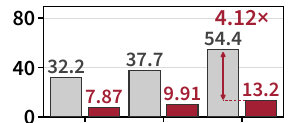}\\
    \includegraphics{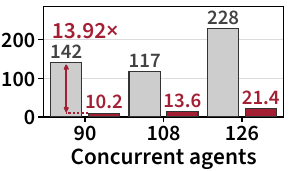}
    \caption{CC (GLM-5.2)}
    \label{fig:eval-single-cc}
  \end{subfigure}\hspace{0.04in}%
  \begin{subfigure}[t]{1.94in}
    \centering
    \captionsetup{margin={0.29in,0.05in}}
    \includegraphics{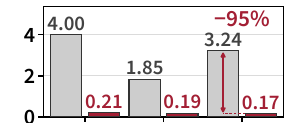}\\
    \includegraphics{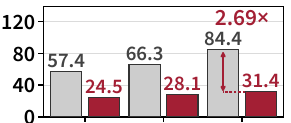}\\
    \includegraphics{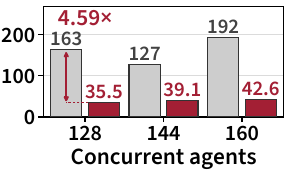}
    \caption{OSWorld (Kimi K2.6)}
    \label{fig:eval-single-osworld}
  \end{subfigure}
  \caption{\textbf{\sys{} loses less KV reuse and achieves lower mean E2E latency
  than Dynamo in every setting.} Lost KV reuse equals $1 - h/h^*$, with $h$ the KV
  hit rate and $h^*$ the hit rate under unlimited KV cache capacity shared by all
  nodes and no phase switches. In each panel, an arrow marks the largest gap, and
  the number above it gives the gap as a relative reduction for lost KV reuse and
  as a ratio for latency.}
  \label{fig:eval-single}
  \vspace{-4mm}
\end{figure}

\section{Experiments}
\label{sec:evaluations}

In this section, we evaluate \sys{} on coding, computer-use, and math agents
served on 36 GB300 GPUs. \autoref{sec:eval-setup} summarizes the setup.
\autoref{sec:eval-single} evaluates each agent workload in isolation under a
stationary load. \autoref{sec:eval-shift} evaluates a shifting workload
composition. \autoref{sec:eval-ablation} measures the sensitivity of \sys{} to
the initial PD split, the contribution of each component, and the cost of online
phase-split adaptation. \appref{app:experiments} further evaluates \sys{} on up
to 256 GPUs, in simulation at 3{,}072 GPUs, under open-loop arrivals, and with
different parallelism for prefill and decode.

\subsection{Experimental setup}
\label{sec:eval-setup}

\textbf{Workloads.} We deploy mini-SWE-agent~\cite{yang2024sweagent} on
PutnamBench~\citep[Putnam;][]{tsoukalas2024putnambench} and Anthropic's
computer-use agent~\cite{anthropic2024computeruse} on
OSWorld~2.0~\citep[OSWorld;][]{yuan2026osworld}. Putnam is a decode-heavy
workload that runs a single agent branch per problem without context
management. OSWorld is a prefill-heavy, multimodal, multi-agent workload that
appends a screenshot to the context at every step
(\autoref{fig:intro-diversity}). The Claude Code
traces~\citep[CC;][]{semianalysis2026cctraces}, collected from real users, form
the most dynamic of the three workloads. They interleave sub-agents,
compactions, and long pauses between requests (\autoref{fig:intro_context}).
Putnam and OSWorld thus differ most in their demand for prefill and decode,
whereas CC exercises agent context management the most. We replay CC with the
Trie replayer~\cite{appliedcompute2026trie}, which preserves the pauses between
requests and the prefix sharing of the original traces. Trie fixes the length
of each output and the duration of each tool call to their recorded values.
Later requests carry these outputs and the recorded tool results, thereby
preserving the prefix that each request shares with earlier requests.

\textbf{Models and deployment.} We serve
GLM-5.2~\cite{glm5team2026glm5vibecodingagentic} for Putnam and CC and Kimi
K2.6~\cite{kimi2026kimik26} for OSWorld and for the run with a shifting workload
composition. Each model is quantized to NVFP4 with an FP8 KV cache and uses
multi-token prediction (MTP) for speculative decoding. It runs with four-way
tensor parallelism (TP4) on nine 4$\times$GB300 nodes~\cite{nvidia2025gb300}. On
each node, host memory (L2) caches the KV of 4$\times$ as many tokens as GPU
memory (L1).

\textbf{Evaluation settings.} \sys{} uses a planning window of
$T_{\mathrm{win}} = 600$~s with $m = 10$ slices and learns its iteration-time
models online (\autoref{sec:design-model}). \appref{app:details-planner} lists
the other parameters of our runs. Each run starts with a 30-minute warmup, which
we exclude from measurement. After the warmup, stationary runs serve one agent
workload for one hour at a fixed concurrency, replacing each finished agent with
a new one. We test three concurrency levels per workload, namely 224, 256, and
288 agents for Putnam, 90, 108, and 126 for CC, and 128, 144, and 160 for
OSWorld. The run with a shifting workload composition serves 288 concurrent
Putnam agents, 144 Putnam and 72 OSWorld agents, and 144 OSWorld agents in three
consecutive hours after the warmup, shifting demand toward prefill.

\begin{figure}[t]
  \centering
  \captionsetup[subfigure]{skip=2pt}
  \includegraphics{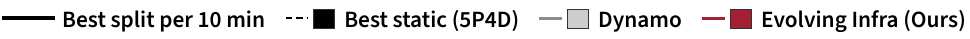}\\[1pt]
  \begin{subfigure}[t]{2.90in}
    \centering
    \captionsetup{margin={0.54in,0.13in}}
    \includegraphics[width=\linewidth]{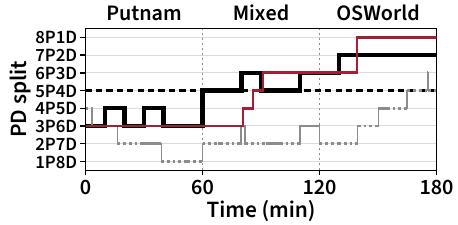}
    \caption{PD split over time}
    \label{fig:eval-shift-split}
  \end{subfigure}\hfill
  \begin{subfigure}[t]{1.69in}
    \centering
    \captionsetup{margin={0.285in,0.05in}}
    \includegraphics[width=\linewidth]{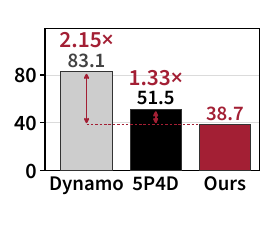}
    \caption{Mean E2E latency (s)}
    \label{fig:eval-shift-e2e}
  \end{subfigure}\hfill
  \begin{subfigure}[t]{1.69in}
    \centering
    \captionsetup{margin={0.285in,0.05in}}
    \includegraphics[width=\linewidth]{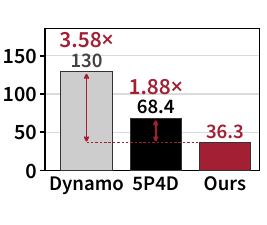}
    \caption{P95 TTFT (s)}
    \label{fig:eval-shift-ttft}
  \end{subfigure}
  \caption{\textbf{\sys{} adapts the PD split as the workload composition shifts
  and achieves lower latency than Dynamo and the best static split.} (a) The best
  split per 10-minute window and the best static split are profiled under SGLang
  PD disaggregation, as in \autoref{fig:intro-dynamicpd}. The Dynamo line is
  dotted while one of its nodes relaunches and only eight nodes serve requests.
  (b, c) Over the whole run, the red number above a baseline bar gives its ratio
  to \sys{}.}
  \label{fig:eval-shift}
\end{figure}

\textbf{Baselines.} We compare \sys{} with Dynamo v1.5.1~\cite{nvidia2025dynamo}
in the configuration of \autoref{sec:motivation}. To our knowledge, Dynamo is
the state-of-the-art open-source system for dynamic PD disaggregation. Both
start from the prefill-heavy split 6P3D, a common assumption for agentic
workloads~\cite{together2026codingagents}. For the shifting workload
composition, the ablation (\autoref{sec:eval-ablation}), and the lowest
concurrency of each stationary workload (\appref{app:exp-static-split}), we
also compare with the best static split, which runs SGLang PD disaggregation at
the fixed split with the highest throughput over the whole
run~\cite{patel2024splitwise}. All baselines use the SGLang Model
Gateway~\cite{sglang2026sglang} with cache-aware prefill routing and round-robin
decode routing, the setup with the lowest P95 TTFT in \autoref{sec:motiv-kv}.
Each node runs HiCache~\cite{xie2025sglang} with LRU eviction and the same
deduplicated host memory as \sys{} (\appref{app:cpu-dedup}). We do not compare
with the runtime role switch of SGLang~\cite{sglang2026roleswitch}, because it
rejects speculative decoding, which both models use.
Llumnix~\cite{llumnix2026adaptivepd} cannot run on the Arm CPUs of our GB300
nodes, and \appref{app:related-split} compares it with \sys{} on H200 GPUs.

\subsection{Stationary workloads}
\label{sec:eval-single}

\begin{figure}[t]
  \centering
  \captionsetup[subfigure]{skip=2pt}
  \makebox[3.6in][l]{\hspace{0.34in}\makebox[3.22in]{\includegraphics{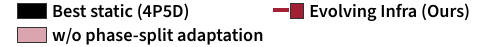}}}\hspace{0.4in}%
  \makebox[1.6in][l]{\hspace{0.34in}\makebox[1.22in]{\includegraphics{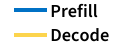}}}\\[-2pt]
  \begin{subfigure}[t]{1.6in}
    \centering
    \captionsetup{margin={0.34in,0.04in}}
    \includegraphics{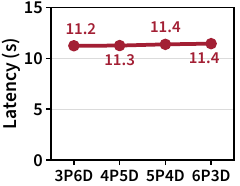}
    \caption{Initial PD split}
    \label{fig:eval-ablation-init}
  \end{subfigure}\hspace{0.4in}%
  \begin{subfigure}[t]{1.6in}
    \centering
    \captionsetup{margin={0.34in,0.04in}}
    \includegraphics{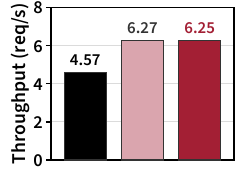}
    \caption{Online adaptation}
    \label{fig:eval-ablation-adaptation}
  \end{subfigure}\hspace{0.4in}%
  \begin{subfigure}[t]{1.6in}
    \centering
    \captionsetup{margin={0.34in,0.04in}}
    \makebox[\linewidth][l]{\includegraphics{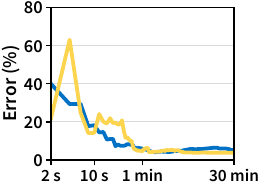}}
    \caption{Model convergence}
    \label{fig:eval-ablation-convergence}
  \end{subfigure}
  \caption{\textbf{\sys{} achieves similar latency from every initial PD split,
  online phase-split adaptation costs almost no throughput, and the
  iteration-time models converge within a minute.} CC with 126 concurrent agents.
  (a)~Mean E2E latency from each initial split. (b)~Throughput at the best static
  split, 4P5D. (c)~Mean error of the iteration-time models against learning
  time.}
  \label{fig:eval-ablation}
  \vspace{-3mm}
\end{figure}

\textbf{Low-cost online adaptation across agent workloads.} \sys{} outperforms
Dynamo in throughput, lost KV reuse, and mean E2E latency for every agent
workload (\autoref{fig:eval-single}). It achieves \textbf{1.10--2.73$\times$}
the throughput of Dynamo (\appref{app:exp-throughput}), reduces lost KV reuse by
\textbf{81--95\%}, and lowers the mean E2E latency by
\textbf{1.15--4.12$\times$} and the P95 TTFT by up to \textbf{13.92$\times$}. At
the lowest concurrency of each workload, it also achieves 0.98--1.38$\times$ the
throughput of the best static split under SGLang
(\appref{app:exp-static-split}). Averaged over all nine settings, a phase switch
of \sys{} completes in \textbf{6.2}~s, including the wait for the running
requests of the node, which the node keeps serving. This is over
\textbf{100}$\times$ shorter than a Dynamo relaunch, which takes 754~s on average
and keeps the node out of service throughout (\appref{app:exp-switch}). The
symmetric engine (\autoref{sec:dataplane-switch}) and the shared KV pool
(\autoref{sec:dataplane-pool}) thus avoid the relaunch and retain the KV of the
switching node, the two costs of a phase switch in \autoref{sec:motiv-switch}.
They stay fast when prefill and decode use different parallelism
(\appref{app:exp-hetero-parallel}).

\textbf{Accurate compute allocation for diverse agent workloads.} \sys{} reduces
the mean E2E latency by \textbf{3.80--4.12$\times$} on CC,
\textbf{2.34--2.69$\times$} on OSWorld, and \textbf{1.15--1.23$\times$} on
Putnam relative to Dynamo (middle row of \autoref{fig:eval-single}). The planner
(\autoref{sec:design-split}) contributes to this gain by matching the PD split
to the compute demand of each workload purely online. It selects 3P6D--4P5D for
CC and 8P1D for OSWorld (\appref{app:exp-split-time}). These splits differ from
the best static splits under SGLang in \autoref{fig:intro-diversity}, 4P5D and
6P3D, since the symmetric engines shift the best split
(\appref{app:exp-static-split}). On the decode-heavy Putnam, the planner keeps
only one or two prefill nodes to maximize throughput. This choice trades P95
TTFT for throughput, raising the P95 TTFT above that of Dynamo at 224 and 288
agents while lowering the mean E2E latency. Unlike the coarse signals in
\autoref{sec:motiv-demand}, the iteration-time models separate cached from
uncached tokens (\autoref{sec:design-model}), and the average work of a request
barely changes during short bursts. \sys{} thus switches only \textbf{2.0} times
per run on average, whereas the Dynamo planner reacts to queue depth and KV
utilization, switches \textbf{3.6} times per run, and deviates from the best
static split of each workload for \textbf{41--100\%} of the measurement period
(\appref{app:exp-split-time}). Under open-loop arrivals, the planner also gives
prefill more nodes as agents arrive faster, and \sys{} sustains twice the
arrival rate of the best static split (\appref{app:exp-open-loop}).

\textbf{Accurate KV cache management under agent context management.} Since
asynchronous fetches make moving KV fast (\autoref{sec:dataplane-pool}), the
router routes each request to the node with the shortest predicted TTFT
(\autoref{sec:design-placement}) rather than to the few nodes holding its prefix
(\autoref{sec:motiv-kv}). \sys{} thus reduces the P95 TTFT by
\textbf{3.25--13.92$\times$} on CC and OSWorld relative to Dynamo (bottom row of
\autoref{fig:eval-single}). On CC, the most loaded prefill node queues requests
\textbf{1.75--1.85$\times$} as long as the average node, versus
\textbf{2.29--2.61$\times$} under cache-aware routing (\appref{app:exp-queue}).
With the counts of the directory (\autoref{sec:dataplane-directory}), the
eviction of \sys{} (\autoref{sec:design-placement}) retains nearly all the KV
that agents still need (top row of \autoref{fig:eval-single}). In contrast, the
LRU eviction of Dynamo drops the KV of branches that wait for tools, sub-agents,
or users, which accounts for \textbf{34--46\%} of its lost KV reuse on CC. The
fraction of prompt tokens that Dynamo prefills without cached KV is up to
\textbf{2.99$\times$} that of \sys{} (\appref{app:exp-switch}).

\subsection{Shifting workload composition}
\label{sec:eval-shift}

\textbf{Precise and timely adaptation to a shifting workload composition.} The PD
split of \sys{} tracks the best split of each 10-minute window as the planner
(\autoref{sec:design-split}) follows the shift from Putnam to OSWorld
(\autoref{fig:eval-shift-split}). In the OSWorld period, it settles at 8P1D, one
prefill node more than the best split under SGLang, since the symmetric engines
shift the best split (\appref{app:exp-static-split}). The best static split
balances prefill and decode only in the mixed period
(\autoref{fig:intro-dynamicpd}). The Dynamo planner keeps fewer prefill nodes
than the best split for most of the run, as \autoref{sec:motiv-demand} predicts.
\sys{} thus reduces the mean E2E latency by \textbf{2.15$\times$} and
\textbf{1.33$\times$} and the P95 TTFT by \textbf{3.58$\times$} and
\textbf{1.88$\times$} relative to Dynamo and the best static split, respectively
(\autoref{fig:eval-shift}b, c). It also achieves \textbf{2.14$\times$} the
throughput of Dynamo and \textbf{1.28$\times$} that of the best static split, and
the highest throughput in every period (\appref{app:exp-throughput}).

\subsection{Ablation study}
\label{sec:eval-ablation}

\textbf{Sensitivity to the initial PD split.} \sys{} achieves a similar mean E2E
latency from every initial split (\autoref{fig:eval-ablation-init}). Serving CC
with 126 concurrent agents from four initial splits, from 3P6D to 6P3D, its mean
E2E latency varies by at most \textbf{1.8\%}. In all four runs, the planner
converges to 4P5D (\appref{app:exp-split-time}). On this workload, online
phase-split adaptation thus does not need a well-chosen initial split.

\textbf{Contribution of each component.} On CC with 126 concurrent agents, the
infrastructure of \autoref{sec:dataplane} and the KV-reuse-aware eviction and
routing of \autoref{sec:design-placement} alone achieve \textbf{1.37$\times$} the
throughput of the best static split at the same 4P5D split, without online
phase-split adaptation (\autoref{fig:eval-ablation-adaptation}). With the routing
of SGLang, asynchronous KV fetching alone achieves \textbf{1.12$\times$} the
throughput of upstream SGLang, because it cuts the idle time of prefill nodes
from 23.6\% to 14.4\% (\appref{app:exp-async-fetch}). KV-reuse-aware routing and
eviction contribute the rest. Compared with cache-aware routing at the same
split, \sys{} loses 0.7\% instead of 2.4\% of the KV reuse and prefills 31\%
fewer uncached prompt tokens per request (\appref{app:exp-thrashing}). Pausing
agents as ThunderAgent does brings no such gain (0.95$\times$), because KV
capacity does not limit serving on GB300 nodes (\appref{app:exp-thrashing}).

\textbf{Cost of online phase-split adaptation.} Enabling online phase-split
adaptation at the best static split lowers the throughput by only
\textbf{0.3\%} (\autoref{fig:eval-ablation-adaptation}). Online adaptation thus
costs almost no throughput even when the initial split is already the best. It
also avoids offline profiling, which takes 10{,}752 GPU-hours for the three-hour
CC trace on 256 GPUs even with a binary search and recurs for every agent
workload (\appref{app:exp-scale}). Online fitting is fast as well. The
iteration-time models converge to a low mean error within a minute of learning
(\autoref{fig:eval-ablation-convergence}), far shorter than the 30-minute warmup
of our runs.
\vspace{-2mm}
\section{Conclusions}
\label{sec:conclusions}
We present \sys{}, an agent-adaptive serving framework that decouples compute
allocation from KV cache management. Existing PD-disaggregated systems tie the
KV cache of each node to the phase of that node. Changing the PD split thus
either relaunches a node or loses its KV. Keeping KV on its node also forces
routing to trade load balance against cache locality. In \sys{}, symmetric
engines switch the phase of a node in place. A shared KV pool keeps the KV of
every node readable by all other nodes.

On this data plane, \sys{} adapts compute allocation to diverse agent workloads
and their shifting composition, and KV cache management to agent context
management. Its planner reassigns nodes between prefill and decode with
iteration-time models that separate cached from uncached tokens and are fitted
online. Its router balances predicted TTFT across prefill nodes. With the branch
counts of the KV directory, each node evicts first the KV that no active branch
references, such as replaced histories and the contexts of finished sub-agents.

On 36 GB300 GPUs, a phase switch of \sys{} takes 6.2~s on average, over
100$\times$ shorter than a Dynamo relaunch. Under stationary load, \sys{}
delivers up to 2.73$\times$ the throughput of Dynamo, loses 81--95\% less KV
reuse, and reduces the mean E2E latency by up to 4.12$\times$. Under a shifting
workload composition, it reduces the mean E2E latency by 2.15$\times$ and
1.33$\times$ relative to Dynamo and the best static split, respectively. These
gains hold on 256 GB300 GPUs and, in simulation, on 3{,}072 GPUs
(\appref{app:exp-scale}).

\bibliographystyle{main}
\bibliography{references}

\clearpage
\appendix
\counterwithin{figure}{section}
\counterwithin{table}{section}
\counterwithin{equation}{section}
\counterwithin{algorithm}{section}
\onecolumn
 
\section{Additional Related Work}
\label{app:related}
 
This appendix extends the review of \autoref{sec:back-existsystem}. \appref{app:related-split}
details how existing systems change the PD split.
\appref{app:related-kvstruct} and \appref{app:related-dynamo} describe the KV
tiers and the KV index that Sections~\ref{sec:dataplane-pool} and~\ref{sec:dataplane-directory} build on.
\appref{app:related-agent} compares agent-aware KV management with \sys{}, and
\appref{app:cpu-dedup} explains why host memory can hold the KV that agents
reuse.
 
\subsection{Prefill--decode split adaptation}
\label{app:related-split}
 
\begin{figure}[b]
  \centering
  \captionsetup[subfigure]{skip=2pt}
  \begin{subfigure}[t]{2.81in}
    \centering
    \captionsetup{margin={0.38in,0.05in}}
    \includegraphics{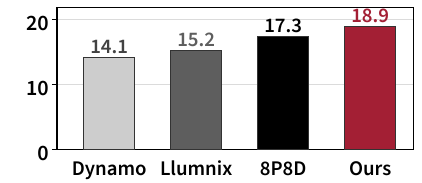}
    \caption{Throughput (req/s)}
    \label{fig:related-split-throughput}
  \end{subfigure}\hspace{0.3in}
  \begin{subfigure}[t]{2.73in}
    \centering
    \captionsetup{margin={0.30in,0.05in}}
    \includegraphics{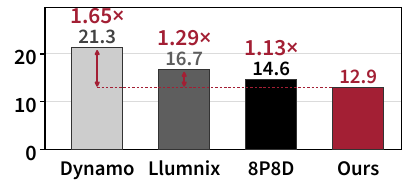}
    \caption{Mean E2E latency (s)}
    \label{fig:related-split-e2e}
  \end{subfigure}
  \caption{\textbf{Dynamo and Llumnix adapt the PD split but fall behind the best static split.} DeepSeek-V4-Flash runs on 16 engines of four H200 GPUs each (TP4) and serves CC at 384 concurrent sessions. Dynamo, Llumnix and \sys{} start from 12P4D. Of the two static splits we ran, 8P8D and 9P7D, 8P8D is the best static split. (a)~Both baselines deliver less throughput than the best static split, which \sys{} exceeds by 9\%. (b)~The red number above a baseline bar gives its ratio to \sys{}.}
  \label{fig:related-split}
\end{figure}
 
Current serving engines fix their phase at launch by default. Systems that adapt the PD split around this limit either relaunch engines, which takes minutes and empties their KV caches, or switch them in place. Those that switch in place lose the KV of a switched engine or size the pools with a coarse cost model. Others keep the roles of nodes fixed and shift only compute or KV between the phases, inside an engine or between fixed prefill and decode nodes.
 
\paragraph{vLLM and SGLang relaunch to change phase.}
In SGLang, a flag at launch makes each engine a prefill or a decode engine, and the router is given separate lists of prefill and decode engines~\cite{sglang2026pd}. The two roles run different scheduling loops with
role-specific state. A prefill engine keeps a bootstrap queue and sends KV to decode engines. A decode engine reserves KV slots for each incoming request
before its prefill finishes, receives KV through its own threads and queues,
and captures CUDA graphs for decoding~\cite{sglang2026pd,sglang2026roleswitch}.
vLLM likewise configures each engine at launch to produce KV for prefill or
to consume it for decode, and its proxy is given fixed lists of prefill and decode engines~\cite{vllm2026nixl}. Changing the phase of an engine thus
requires relaunching it with a new configuration. SGLang's recent opt-in
runtime switch avoids the relaunch, but it still drains the engine to idle and
releases its prefix cache~\cite{sglang2026roleswitch}.
Systems built on these engines therefore
relaunch them. The Dynamo planner changes the phase of a node by terminating its worker, without waiting for in-flight requests, and launching a worker of the other phase~\cite{nvidia2026planner}. The new worker must load the model and
starts with an empty KV cache, and Dynamo notes that spinning up a GPU worker
takes minutes~\cite{nvidia2026planner}. In the runs of \autoref{sec:eval-single}, a Dynamo switch takes 754~s on average, from the planner decision until the relaunched worker is ready, and on CC with 108 agents Dynamo loses 4.9\% of the available KV reuse over its four switches (\appref{app:exp-switch}). While a node relaunches, only eight of the nine nodes serve requests, which covers 44--91\% of the one-hour measurement period (\appref{app:exp-split-time}).
 
\paragraph{Arrow switches in place but loses KV.}
Arrow~\cite{wu2025arrow} keeps nodes stateless, so that any node can serve both prefill and decode requests, and it moves nodes between the prefill and decode pools without a restart. Its code is not public, so
\autoref{sec:motivation} evaluates our reimplementation of its planner and scheduler on the
symmetric engines of \sys{}. Arrow has three problems on agentic workloads.
It sizes the pools by TTFT and TPOT, which agent context management makes
fluctuate. It predicts TTFT from the full input length, which ignores KV
reuse. Its scheduler also ignores where KV is cached, and each node keeps its KV in its own cache, so later prefills do not reuse the KV of a node that moves to the decode pool. In our runs on CC, Arrow switches nodes 34 times, ends at 8P1D, and delivers 65\% less throughput than the best static split under SGLang PD disaggregation (\autoref{sec:motiv-demand}). In these runs, it loses 55.5\% of the available KV reuse (\autoref{sec:motiv-switch}).
 
\paragraph{xLLM cannot switch phases and route by cached KV at once.} xLLM~\cite{liu2025xllm} switches nodes between phases in place, but its open-source service implements this switching and KV-aware routing as alternative policies chosen at launch~\cite{xllm2026service}. Only its SLO-aware policy switches phases, and it flips single nodes in reaction to individual requests instead of sizing the split with a planner. This policy predicts TTFT from the full prompt length, so it counts cached prefixes as prefill work and ignores where KV is cached. Its cache-aware policy routes by prefix hits but never switches phases.
 
\paragraph{Llumnix adapts the split with a coarse cost model.} Llumnix~\cite{sun2024llumnix} reschedules running requests across engines by live-migrating each request together with its KV, which balances load, reduces memory fragmentation, and isolates high-priority requests. Its adaptive PD mode assigns phases to engines at runtime to meet TTFT and TPOT targets~\cite{llumnix2026adaptivepd}, using a cost model that predicts TTFT by interpolating between profiled points by the number of prefill tokens and the decode load~\cite{llumnix2026slo}. A single token count cannot separate the uncached tokens of an agent request from the long cached prefix they attend to. Part of its KV transfer stack is also released only as a pre-built binary without source code, so we could not run Llumnix on the Arm CPUs of our GB300 nodes. In our runs on H200 GPUs, Llumnix delivers 12.2\% less throughput than the best static split (\autoref{fig:related-split}).
 
\paragraph{semi-PD shares KV between the phases only inside an engine.} semi-PD~\cite{hong2025semipd} runs a prefill process and a decode process on every GPU of an engine, divides the SMs of each GPU between them through NVIDIA MPS, and keeps one copy of the weights and the KV, its unified storage, in the GPU memory of the engine. Shifting SMs restarts only these processes, while a resident process keeps the weights and the KV, so semi-PD avoids costly phase switches inside an engine. Its controller, however, shifts SMs to the phase whose TTFT or TPOT misses its target, and the prefill bursts of agents, such as sub-agents that start together, perturb both (\autoref{sec:motiv-demand}). Across engines, its router weighs the KV cache hit rate of each engine against GPU utilization and queue depth, the trade-off between cache locality and load balance of \autoref{sec:motiv-kv}.
 
\paragraph{MemServe and PPD reuse multi-turn KV but keep the split fixed.} MemServe~\cite{hu2024memserve} and PPD~\cite{li2026prefills} let the later turns of a conversation reuse the KV that its previous turn left on the decode side. MemServe moves this KV to the prefill engine: the decode engine sends the KV that it produced back after each request, and a global scheduler routes each request by where its prefix is cached. PPD instead moves the prefill to the KV: it prefills the uncached tokens of a later turn on the decode node that holds the KV of the conversation, which slows decoding far less than a full prefill does, and a table profiled offline decides per request whether to take this path. Neither, however, adapts the split: MemServe's engines are prefill-only, decode-only, or colocated, and PPD restarts its engines to change the split, so neither follows the shifting workload composition (\autoref{fig:intro-dynamicpd}). PPD's offline table, as its authors note, becomes suboptimal when the workload drifts far from the one it was profiled on. MemServe also judges KV reuse from stale records: its scheduler learns where KV resides only from the responses that pass through it, so it misses evictions and expires its records after a fixed time-to-live. PPD keeps the KV of a conversation on one decode node, so each later turn either runs on that node or recomputes the previous response on a prefill node.
 
\sys{} instead switches the phase of a node in place, and the shared KV pool
keeps the KV of the node readable by all other nodes (Sections~\ref{sec:dataplane-switch}
and~\ref{sec:dataplane-pool}).
 
\subsection{KV memory hierarchy in current serving engines}
\label{app:related-kvstruct}
 
\autoref{sec:dataplane-pool} builds the shared KV pool of \sys{} on SGLang HiCache. This
subsection describes how current engines tier KV, how they share it across
nodes through Mooncake, and why this sharing adds latency.
 
\paragraph{Private host tiers.}
SGLang HiCache backs the GPU KV cache (L1) of an engine with host memory (L2)
and an optional storage backend (L3) such as Mooncake
Store~\cite{xie2025sglang,sglang2026hicache}. L1 and L2 are private to each
engine, and every tensor-parallel rank allocates its own L2 pool. Engines
therefore never read each other's L2: KV that one engine computes becomes
visible to another only after it reaches L3~\cite{sglang2026hicache}.
 
\paragraph{Sharing through Mooncake.}
Mooncake Store pools the DRAM that nodes contribute into a cluster-wide object
store. A master service keeps the metadata of each object, and clients move the
data directly over RDMA with the Mooncake Transfer
Engine~\cite{mooncake2026store,mooncake2026transfer}. A put first asks the
master to allocate space and then commits the object, and a get first asks the
master where the object is stored. SGLang backs up pages from L2 to Mooncake in
the background, keyed by a hash of the prefix that ends at each page. When a
request is queued, an engine asks Mooncake which pages of its prefix exist,
prefetches them into its own L2, and loads them into GPU memory layer by layer during prefill~\cite{sglang2026hicache,sglang2026source}. vLLM's Mooncake store connector
instead registers GPU memory with Mooncake and moves KV blocks directly between GPU memory and the pool~\cite{vllm2026mooncake}, and a newer SGLang option links its radix cache to Mooncake in the same way, without HiCache's host tier~\cite{sglang2026source}.
 
\paragraph{Why fetching through the pool adds latency.}
KV takes extra hops. With HiCache, it moves from the L2 of the producing engine into a Mooncake segment, which may lie on a third node, and from there into the L2 of the consuming engine before it reaches the GPU. Every lookup, put, and get also involves RPCs to the master, and a page becomes visible only after its backup commits~\cite{mooncake2026store}. The prefetch starts when the request is queued but fills only L2, and the GPU loads the pages layer by layer during prefill. If the prefetch has not finished when the prefill could start, SGLang stops it under the best-effort policy and recomputes the pages that have not arrived, waits for it under the wait-complete policy, and under the default timeout policy waits until a deadline and then recomputes the rest~\cite{sglang2026hicache,sglang2026source}. Mooncake's own benchmark accordingly finds HiCache slightly faster with L2 alone than with Mooncake at equal hit rates in its first rounds, because L2 alone avoids fetching from remote storage~\cite{mooncake2026hicachebench}. The same KV can also occupy the L2 of the producer, the pool, and the L2 of each consumer at once. \sys{} instead lets prefill nodes read KV directly from the L2 of any peer over RDMA, restores it into GPU memory before the prefill begins, and keeps L2 across phase switches (Sections~\ref{sec:dataplane-switch} and~\ref{sec:dataplane-pool}).
 
\subsection{Dynamo KV indexer and router}
\label{app:related-dynamo}
 
The KV directory of \sys{} (\autoref{sec:dataplane-directory}) follows the KV indexer of the Dynamo router. This subsection describes how the index is built and updated, how the router scores nodes with it, and what it does not record.
 
\paragraph{Cluster-wide KV index.}
The KV router of Dynamo keeps a cluster-wide index of cached KV in a prefix tree~\cite{nvidia2026routerdesign}. Each entry of the tree stands for one KV block, keyed by a hash of its tokens, and records the nodes that hold the block. To route a request, the router walks the tree along the blocks of the request and finds, for every node, how many leading blocks it already holds.
 
\paragraph{KV events.}
Engines keep the tree current through KV events. SGLang and vLLM publish an event over ZMQ whenever they store or evict blocks, and a publisher on each node relays the events to the routers~\cite{nvidia2026kvevents}. The engine identifies each block by a prefix hash over all tokens up to and including the block. When it stores new blocks, its event carries their tokens and prefix hashes and the prefix hash of their parent block. Through a per-node table from prefix hashes to tree entries, the router finds the entry of the parent block, descends from it along the token hashes of the new blocks, creating entries as needed, and adds the node to each of these entries~\cite{nvidia2026kvevents,nvidia2026routerdesign,nvidia2025dynamo}. When the engine evicts blocks, its event names them by their prefix hashes, and the router removes the node from their entries. With HiCache, SGLang also names the memory tier in each event, and the router tracks host memory and disk in separate indexes beside the GPU tree~\cite{nvidia2026offloading,nvidia2025dynamo}.
 
\paragraph{Scoring.}
For each candidate node, the router adds the prompt blocks of the request to the prefill blocks already assigned to the node, subtracts a credit for each leading block that the node already holds, and adds the blocks that the node would be decoding. The router tracks these loads itself from the requests that it routes, not from KV events, and sends the request to the node with the lowest cost~\cite{nvidia2026routerdesign}. A block held in GPU memory earns a full credit, while blocks in host memory and on disk earn 0.75 and 0.25 by default~\cite{nvidia2026offloading}. The host and disk credits are constants: they depend neither on how long loading the blocks takes nor on the running requests that the loading delays on the node. \sys{} instead predicts the TTFT of the request on each prefill node, including the part of a KV fetch that the wait in the queue does not hide (\autoref{sec:design-placement} and \appref{app:details-planner}).
 
\paragraph{What the index does not record.}
The index records where each block is cached, on which nodes and in which tier, but not which requests or agent branches will reuse it. KV events report only stores and evictions, not the forks, compactions, and finished sub-agents that decide this reuse. Eviction therefore stays with each engine, and SGLang evicts in LRU order by default~\cite{sglang2026radix}. Dynamo can also pass a per-request priority from the harness to SGLang, which can then evict the KV of the lowest-priority requests first~\cite{nvidia2026sglang}. Neither recency nor priority changes when compaction replaces a history or a KV transfer leaves a copy behind: these blocks keep their last access time, and a cached prefix keeps the highest priority among the requests that reached it~\cite{sglang2026radix}. \sys{} keeps the same event-driven tree and adds one attribute to each block, the number of active branches that reference it on each node (\autoref{sec:dataplane-directory}). The counts follow agent context management and KV transfers, and each node evicts the host-memory blocks with zero counts first (\autoref{sec:design-placement}).
 
\subsection{Agent-aware KV management and routers}
\label{app:related-agent}
 
This subsection compares \sys{} with systems that use the structure of agents
to manage KV.
 
\paragraph{Program-aware scheduling.}
ThunderAgent~\cite{kang2026thunderagent} and Continuum~\cite{li2025continuum}
schedule the requests of an agent together so that its KV survives tool calls.
Continuum pins the KV of a multi-turn program in GPU memory for a time-to-live
that it estimates from the durations of past calls to the same tool.
ThunderAgent abstracts each agent workflow as a program. When the programs on a
data-parallel replica need more KV than its GPU cache holds, ThunderAgent
pauses programs that wait for tools, evicts their KV, and resumes them later,
possibly on another replica with free memory. Both systems thus target
KV-cache thrashing, in which the KV of a waiting program is evicted from GPU
memory and its whole context must be prefilled again. Both run prefill and
decode on the same nodes, and both treat the context of a program as growing
only by appending.
 
\paragraph{Thrashing on current models and hardware.}
Thrashing occurs only when programs need more KV than a replica can hold. Current
engines back GPU memory with a host-memory tier, and compact KV representations
together with deduplication let this tier hold several times as much KV as GPU
memory (\appref{app:cpu-dedup}). A context evicted from GPU memory then stays
in host memory and is reloaded rather than recomputed. ThunderAgent found such offloading too slow over PCIe~\cite{kang2026thunderagent}, but
on GB300 nodes the reload runs over the NVLink-C2C link between the Grace CPUs
and the GPUs~\cite{nvidia2025gb300,aubrey2025inside}. In our runs, Dynamo, whose SGLang engines
evict host memory in LRU order, loses at most 7.8\% of the available KV reuse,
and \sys{} at most 1.0\%, without pausing any agent (\appref{app:exp-switch}). Host memory instead keeps KV that no active branch needs: the source copies that
KV transfers leave, the histories that compaction replaces, and the contexts of
finished sub-agents. \sys{} therefore counts the branches that reference each
copy (\autoref{sec:dataplane-directory}) instead of pausing agents. \appref{app:exp-thrashing}
compares \sys{} with ThunderAgent and cache-aware routing on GB300 nodes.
 
\paragraph{Agent-aware KV management.}
KVTether~\cite{fu2026capture} exports runtime context operations from agent harnesses and maps them to KV lifecycle states, accounting for shared prefixes across contexts. These states guide cache admission, reclamation, and eviction in a multi-tier cache with cross-node transfers. Its presented design focuses on KV lifecycle management and does not address online adaptation of the PD split; its evaluation uses a fixed PD configuration.
 
\subsection{Extending host memory with deduplication}
\label{app:cpu-dedup}
 
\autoref{sec:dataplane-pool} keeps the KV that agents reuse in host memory (L2), and
\appref{app:related-agent} argues that this tier largely absorbs KV thrashing.
This subsection explains why L2 is large enough.
 
\paragraph{Compact KV.}
Although smaller than L3, L2 is sufficient for recent open
models~\citep{glm5team2026glm5vibecodingagentic,kimi2026kimik25}. These models increasingly adopt
compact KV representations such as MQA~\cite{shazeer2019fast} and
MLA~\cite{liu2024deepseekv2}. In MLA, all attention heads share one latent
vector of 576 values per token and layer. With an FP8 KV cache, a token takes
about 35~KB of KV in Kimi K2.6 and about 50~KB in GLM-5.2, including the keys
of the sparse-attention indexer of GLM-5.2. Full multi-head attention at the
same precision would need 36$\times$ and about 50$\times$ as much.
 
\paragraph{Deduplication across tensor-parallel ranks.}
Because all heads share the latent vector, tensor parallelism does not split
MLA KV: every rank holds the complete KV of each token~\cite{sglang2026hicache}.
SGLang HiCache allocates its host pool per rank and thus stores the same KV
once per GPU in L2, although only one rank writes it to the storage
backend~\cite{sglang2026hicache}. vLLM's CPU offloading instead keeps one copy
of each replicated MLA block in a host region that all ranks share, so that at
a tensor-parallel degree of $N$ the host tier holds about $N$ times as many
blocks~\cite{vllm2026mladedup}. LMCache saves MLA KV from one rank and
broadcasts it to the other ranks on retrieval~\cite{lmcache2025mla}. Our
engines deduplicate L2 in the same way: L2 stores a single copy per node for
all attention heads and broadcasts it to all GPUs of the node when fetched in
L1. All systems in our evaluation, including the baselines, use this
deduplication. When serving GLM-5.2 on NVIDIA GB300, L2 typically holds
\textbf{4}$\times$ as much KV as L1. A GB300 node has about 1~TB of host
memory~\cite{nvidia2025gb300}, which could hold about 20 million tokens of
GLM-5.2 KV as a single copy, but only about 5 million if each of its four GPUs
kept its own copy.
 
Production traces of a large cloud provider show that the cache size needed
for an ideal hit ratio is moderate~\cite{wang2025kvcache}. With a deduplicated
host tier, capacity thus rarely forces recomputation in our runs
(\appref{app:related-agent}), and the remaining waste lies in KV that no
active branch references (\autoref{sec:dataplane-directory}).
 
\section{Extended Theoretical Analysis}
\label{app:theory}
 
This appendix extends the iteration-time models of
\autoref{sec:design-model}. \appref{app:theory-h200} fits both models on
another GPU and model. \appref{app:theory-attention} explains why the prefill
iteration-time model holds for attention architectures other than full
attention. \appref{app:theory-moe} derives $\phi(B)$, the fraction of routed
experts that a decode iteration reads, and \appref{app:theory-context} explains
why the decode iteration-time model omits the context length.
 
\subsection{Iteration-time models on H200 with DeepSeek-V4-Flash}
\label{app:theory-h200}
 
\begin{figure}[htb]
  \centering
  \captionsetup[sub]{justification=centering}
  \begin{subfigure}[t]{0.228\textwidth}
    \centering
    \includegraphics[width=\linewidth]{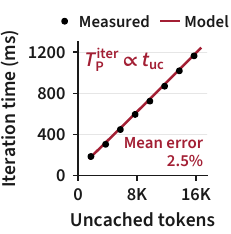}
    \captionsetup{margin={0.10in,0in}}  %
    \caption{Prefill iteration time\\ vs.\ uncached tokens}
    \label{fig:itmodel-h200-uncached}
  \end{subfigure}\hfill
  \begin{subfigure}[t]{0.228\textwidth}
    \centering
    \includegraphics[width=\linewidth]{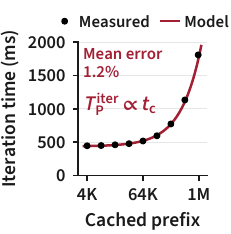}
    \captionsetup{margin={0.10in,0in}}
    \caption{Prefill iteration time\\ vs.\ cached prefix}
    \label{fig:itmodel-h200-prefix}
  \end{subfigure}\hfill
  \begin{subfigure}[t]{0.228\textwidth}
    \centering
    \includegraphics[width=\linewidth]{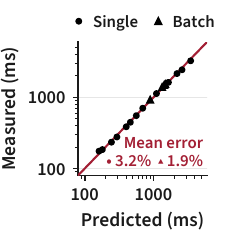}
    \captionsetup{margin={0.10in,0in}}
    \caption{Predicted vs.\\ measured iteration time}
    \label{fig:itmodel-h200-all}
  \end{subfigure}\hfill
  \begin{subfigure}[t]{0.295\textwidth}
    \centering
    \includegraphics[width=\linewidth]{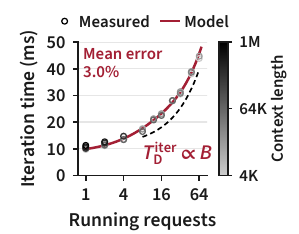}
    \captionsetup{margin={0in,0.05in}}
    \caption{Decode iteration time\\ vs.\ running requests}
    \label{fig:itmodel-h200-decode}
  \end{subfigure}
  \caption{\textbf{The iteration-time models also closely track the iteration times of
  DeepSeek-V4-Flash on H200.} Points are GPU times per iteration of DeepSeek-V4-Flash on four
  H200 GPUs, and red lines are the models in Equations~\eqref{eq:prefill-step}
  and~\eqref{eq:decode-model}, fit on the same settings as in \autoref{fig:itmodel}. Each panel
  reports the mean error on the remaining points. (a)~The prefill iteration time grows nearly
  linearly with the uncached prefill tokens at a 128K cached prefix. (b)~It grows linearly with
  the cached prefix at 8K uncached prefill tokens (log-scale axis). (c)~Outside the fitted
  settings, the mean error is 2.6\% for single requests and 1.9\% for batches of 2--8 requests.
  The panel shows 17 of the 60 measured prefill iterations, evenly spaced and including the four
  batches with the largest errors. (d)~The decode iteration time follows the model as the running
  requests grow, and with 8 or more running requests, the context length (darker for longer)
  changes it by at most 6.9\%.}
  \label{fig:itmodel-h200}
\end{figure}
 
We repeat the measurement of \autoref{fig:itmodel} on a second GPU and model to
test whether the form of the iteration-time models holds beyond GB300 and
GLM-5.2. The engine serves DeepSeek-V4-Flash~\cite{deepseekai2026deepseekv4} on
four H200 GPUs with four-way tensor parallelism (TP4).
DeepSeek-V4-Flash differs from GLM-5.2 in the two components that shape the
models, attention and the MoE layers. Its attention combines a sliding window with compressed attention
(\appref{app:theory-attention}), and its MoE layers route each token to 6 of
256 experts instead of 8 of 256. Its speculative decoding feeds
$t_{\mathrm{fwd}} = 6$ tokens of each running request into every decode
iteration. We measure the same settings as in \autoref{fig:itmodel}. As there, we fit the
prefill model on 20 single-request settings and the decode model on the ten
points with a 16K context.
 
\autoref{fig:itmodel-h200} shows that both models track the measured iteration
times as closely as on GB300. Outside the 20 fitted settings, the prefill model
has a mean error of 2.6\% on the 28 single-request settings and 1.9\% on the 12
batches of 2--8 requests. No batch deviates by more than 4.1\%, including the
batches of eight requests at a 256K cached prefix. Outside the ten fitted
points, the decode model has a mean error of 3.6\%.
 
\subsection{Prefill iteration-time model beyond full attention}
\label{app:theory-attention}
 
Under full attention, each uncached token attends to all cached tokens and to
the uncached tokens before it. A request with $t_{\mathrm{c}}$ cached tokens
and $t_{\mathrm{uc}}$ uncached tokens thus computes
$A = t_{\mathrm{uc}}t_{\mathrm{c}} + t_{\mathrm{uc}}(t_{\mathrm{uc}}+1)/2$
query--key pairs (\autoref{sec:design-model}). Many current models let an
uncached token attend to only part of the earlier tokens. This subsection
explains how these attention layers work and why the prefill iteration-time
model in Equation~\eqref{eq:prefill-step} still expresses their time.
 
\paragraph{Sliding-window attention.} A sliding window lets each token attend
only to a fixed number of the most recent
tokens~\cite{beltagy2020longformer,jiang2023mistral}. Once the context is longer
than the window, every uncached token computes the same number of query--key
pairs, independent of $t_{\mathrm{c}}$. The attention time of a request thus
grows in proportion to $t_{\mathrm{uc}}$, and the term $c_{n,1}\sum
t_{\mathrm{uc}}$ in Equation~\eqref{eq:prefill-step} captures it.
 
\paragraph{Linear attention.} Linear attention replaces the query--key pairs
with earlier tokens by a state of fixed
size~\cite{katharopoulos2020transformers,yang2024gated}. Each uncached token
updates this state and computes its output from it, and the kernels process the
uncached tokens in chunks of a fixed length. Every uncached token thus costs the
same, independent of $t_{\mathrm{c}}$. The attention time of a request grows in
proportion to $t_{\mathrm{uc}}$, and the term $c_{n,1}\sum t_{\mathrm{uc}}$ in
Equation~\eqref{eq:prefill-step} captures it.
 
\paragraph{DeepSeek Sparse Attention (DSA).} DSA, which GLM-5.2
uses~\cite{deepseekai2025deepseekv32,glm5team2026glm5vibecodingagentic}, splits
attention into two steps. A lightweight indexer first scores every earlier token
of each uncached token, and the core attention then attends only to the 2048
tokens with the highest scores. Each score is one query--key pair, and the
indexer thus computes $A$ pairs per request, as full attention does, at a lower
cost per pair. Its time grows with $A$, and the term $c_{n,2}\sum A$ in
Equation~\eqref{eq:prefill-step} captures it. The core attention computes at
most 2048 pairs per uncached token, and its time grows with $t_{\mathrm{uc}}$,
which the term $c_{n,1}\sum t_{\mathrm{uc}}$ captures.
 
\paragraph{Compressed attention in DeepSeek-V4.} Most layers of
DeepSeek-V4~\cite{deepseekai2026deepseekv4} combine a sliding window of 128
tokens with one of two compressed attentions. Both compress the keys and values
of consecutive tokens into entries, and an uncached token attends to these
entries instead of the tokens. Compressed Sparse Attention (CSA) builds one entry
per 4 tokens. Like DSA, its indexer scores all earlier entries, and its core
attention attends to the 512 entries with the highest scores. Heavily Compressed
Attention (HCA) builds one entry per 128 tokens and attends to all earlier
entries. DeepSeek-V4-Flash alternates CSA and HCA in 41 of its 43 layers and
uses the sliding window alone in the other two. The CSA indexer and HCA compute
one pair per 4 and per 128 earlier tokens, about $A/4$ and $A/128$ pairs per
request, and their time grows with $A$. The sliding window, the CSA core
attention, and the compression take a fixed time per uncached token, and their
time per request thus grows with $t_{\mathrm{uc}}$. The terms $c_{n,2}\sum A$
and $c_{n,1}\sum t_{\mathrm{uc}}$ in Equation~\eqref{eq:prefill-step} thus
capture both parts.
 
\subsection{MoE weights in the decode iteration-time model}
\label{app:theory-moe}
 
In an MoE model, a decode iteration reads the dense weights in full but reads a
routed expert only when some token selects it. The dense weights, such as the
attention projections and the shared experts, serve every token, and reading
them is part of the fixed cost $d_{n,0}$ in Equation~\eqref{eq:decode-model}. In each
MoE layer, a gate assigns every token to $k$ of the $E$ routed experts. The
layer then reads the weights of each selected expert from GPU memory once and
applies them to all tokens that select this expert. It does not read the
weights of experts that no token selects. Our engines split every expert across the GPUs of a node with tensor
parallelism, and each GPU reads its share of every selected expert. The time to
read the routed experts thus grows in proportion to the number of selected
experts. Equation~\eqref{eq:decode-model} writes this time as $d_{n,1}\phi(B)$,
where $d_{n,1}$ is the time to read all routed experts and $\phi(B)$ is the
fraction that the iteration selects.
 
We derive $\phi(B)$ under two assumptions. First, every expert receives the
same share of tokens. MoE models are trained to keep the load of their experts
balanced~\cite{deepseekai2024deepseek}, and each token thus selects a given
expert with probability $k/E$. Second, different
tokens select their experts independently. Each of the $B$ running requests
feeds $t_{\mathrm{fwd}}$ tokens into every MoE layer, and the layer thus routes
$t_{\mathrm{fwd}}B$ tokens. The probability that none of them selects a given expert
is $(1-k/E)^{t_{\mathrm{fwd}}B}$. The expected fraction of selected experts is one
minus this probability, which gives $\phi(B) = 1 - (1-k/E)^{t_{\mathrm{fwd}}B}$
in \autoref{sec:design-model}.
 
\subsection{Context length in the decode iteration-time model}
\label{app:theory-context}
 
The decode iteration-time model omits the KV that an iteration reads for its
running requests. This KV grows with their total context but stays small next
to the weights that the iteration reads, for two reasons. First, recent models
store little KV per token. With MLA, a token of GLM-5.2 takes about 50~KB of KV
in FP8, about 50$\times$ less than under full multi-head attention
(\appref{app:cpu-dedup}). Second, sparse attention lets decode read only part
of this KV. GLM-5.2 reads the full KV of at most 2048 selected tokens per layer
and only the small indexer keys of all other tokens. DeepSeek-V4-Flash reads
the full KV of its 128-token window and 512 selected entries, and of all other
tokens only compressed entries or indexer keys that each cover 4 or 128 tokens
(\appref{app:theory-attention}). As a result, one million more context tokens
lengthen a decode iteration by about 1~ms for GLM-5.2 on GB300 and by at most
1.5~ms for DeepSeek-V4-Flash on H200, whereas an iteration with 8 to 64 running
requests takes 17 to 45~ms (\autoref{fig:itmodel-decode} and
\autoref{fig:itmodel-h200-decode}).
 
\section{Extended Analysis of Stationary Workloads and the Shifting Workload Composition}
\label{app:exp-stationary}
 
This appendix explains the gains of \sys{} on the stationary workloads of \autoref{sec:eval-single} and under the shifting workload composition of \autoref{sec:eval-shift}. For the stationary workloads, \appref{app:exp-switch} compares the phase switches and KV reuse of \sys{} and Dynamo, \appref{app:exp-split-time} shows how the PD split of each system changes over time, \appref{app:exp-queue} compares how evenly they balance prefill load, and \appref{app:exp-throughput} relates their throughput to their E2E latency. For the shifting workload composition, \appref{app:exp-throughput} also compares the throughput of \sys{}, Dynamo, and the best static split in each period. The Dynamo baseline in \autoref{sec:evaluations} and Appendices~\ref{app:exp-stationary} and~\ref{app:experiments} corresponds to release v1.5.1~\cite{nvidia2025dynamo}.
 
\subsection{Phase Switches and KV Reuse}
\label{app:exp-switch}
 
\autoref{tab:app-switch} lists, for every setting of \autoref{fig:eval-single},
the phase switches of each system in the measurement period, the mean time per
switch, and the lost KV reuse. For Dynamo, the switch time runs from the planner
decision to the point at which the relaunched engine is ready. For \sys{}, it
runs from the planner decision to the point at which the node serves requests of
its new phase, including the time to drain the requests running on the node. A
phase switch of \sys{} takes \textbf{6.2}~s on average, over \textbf{100$\times$}
shorter than the \textbf{754}~s that Dynamo needs to relaunch an engine, and
retains the KV of the switching node. \sys{} thus loses \textbf{81--95\%} less
KV reuse than Dynamo in every setting, even on Putnam, where it switches more
often. As a result, the fraction of prompt tokens that Dynamo prefills without
cached KV, one minus its KV hit rate in \autoref{tab:app-switch}, is
1.29--2.99$\times$ that of \sys{}.
 
On CC, the KV reuse that Dynamo loses has two main causes
(\autoref{sec:motiv-kv}). We attribute a loss to eviction when the request
runs on the same prefill node as the previous request in its branch, and to
routing when it runs on a different one. Eviction accounts for 39\%, 34\%, and 46\% of the lost KV reuse with 90, 108, and 126 agents, because LRU eviction removes the KV of branches that pause for tools, sub-agents, or users. Routing accounts for another 22\%, 17\%, and 15\%.
 
\begin{table}[ht]
\centering
\small
\caption{\textbf{Phase switches and lost KV reuse on stationary workloads.}}
\label{tab:app-switch}
\setlength{\tabcolsep}{3.5pt}%
\resizebox{\textwidth}{!}{%
\begin{tabular}{@{}llcccccccc@{}}
\toprule
 & & \multicolumn{4}{c}{Dynamo} & \multicolumn{4}{c}{\sys{}} \\
\cmidrule(lr){3-6} \cmidrule(l){7-10}
 & Concurrent & & Mean switch & KV hit & Lost KV & & Mean switch & KV hit & Lost KV \\
Workload & agents & Switches & time (s) & rate (\%) & reuse (\%) & Switches & time (s) & rate (\%) & reuse (\%) \\
\midrule
Putnam & 224 & 2 & 798 & 96.59 & 2.47 & 3 & 2.3 & 98.54 & 0.48 \\
 & 256 & 2 & 798 & 96.56 & 2.48 & 2 & 3.5 & 98.85 & 0.17 \\
 & 288 & 3 & 800 & 96.42 & 2.63 & 5 & 3.4 & 98.56 & 0.46 \\
\midrule
CC & 90 & 4 & 681 & 92.25 & 5.13 & 1 & 0.7 & 97.08 & 0.32 \\
 & 108 & 4 & 831 & 92.62 & 4.89 & 2 & 1.9 & 96.87 & 0.48 \\
 & 126 & 4 & 813 & 89.52 & 7.78 & 2 & 10.8 & 96.25 & 1.01 \\
\midrule
OSWorld & 128 & 5 & 666 & 89.79 & 4.00 & 1 & 19.4 & 93.79 & 0.21 \\
 & 144 & 4 & 730 & 92.01 & 1.85 & 1 & 18.0 & 93.79 & 0.19 \\
 & 160 & 4 & 748 & 90.67 & 3.24 & 1 & 17.7 & 93.84 & 0.17 \\
\bottomrule
\end{tabular}}
\end{table}
 
\subsection{PD Split over Time}
\label{app:exp-split-time}
 
\begin{figure}[ht]
  \centering
  \captionsetup[subfigure]{skip=2pt}
  \includegraphics{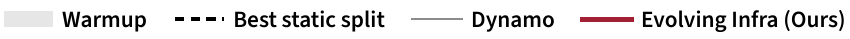}\\[1pt]
  \begin{subfigure}[t]{2.44in}
    \centering
    \captionsetup{margin={0.62in,0.10in}}
    \includegraphics{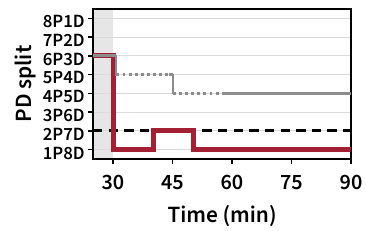}
    \caption{Putnam, 224 agents}
    \label{fig:app-split-putnam}
  \end{subfigure}\hfill
  \begin{subfigure}[t]{1.92in}
    \centering
    \captionsetup{margin={0.10in,0.10in}}
    \includegraphics{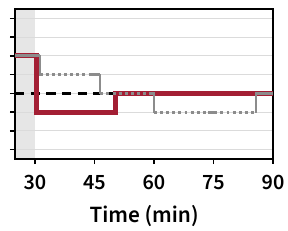}
    \caption{CC, 126 agents}
    \label{fig:app-split-cc}
  \end{subfigure}\hfill
  \begin{subfigure}[t]{1.92in}
    \centering
    \captionsetup{margin={0.10in,0.10in}}
    \includegraphics{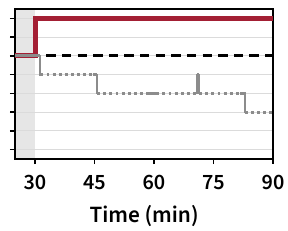}
    \caption{OSWorld, 160 agents}
    \label{fig:app-split-osworld}
  \end{subfigure}
  \caption{\textbf{PD split of \sys{} and Dynamo over time.} The Dynamo line is dotted while one of its nodes relaunches and only eight nodes serve requests.}
  \label{fig:app-split}
  \label{fig:app-split-ours}
  \label{fig:app-split-dynamo}
\end{figure}

\autoref{fig:app-split} shows the PD split of both systems over the run for one
setting of each agent workload. \sys{} changes its split at the first planner
decision after warmup. It moves OSWorld to 8P1D and holds this split, moves CC to
3P6D and later to 4P5D, and alternates between 1P8D and 2P7D on Putnam. On
OSWorld and CC, 8P1D and 3P6D are the best static splits on the symmetric engines
of \sys{} (\appref{app:exp-static-split}). Dynamo changes its split by one node
per relaunch. Each relaunch takes \textbf{10--15} minutes, during which only
eight nodes serve requests, and these relaunches cover \textbf{44--91\%} of the
measurement period, depending on the setting. Across all nine settings of
\autoref{fig:eval-single}, Dynamo deviates from the best static split of each
workload under SGLang PD disaggregation for \textbf{41--100\%} of the
measurement period.
 
\subsection{Prefill Load Balance}
\label{app:exp-queue}
 
\autoref{tab:app-queue} reports the mean time that a prefill request waits in
the queue of a prefill node and, as a measure of load balance, the queueing time
on the most loaded prefill node relative to the average across prefill nodes,
computed per minute and averaged over the measurement period. On CC, \sys{}
keeps this ratio at \textbf{1.75--1.85$\times$}, compared with \textbf{2.29--2.61$\times$} under
Dynamo, and reduces the queueing time by \textbf{2.6--7.6$\times$}.
 
\begin{table}[ht]
\centering
\small
\caption{\textbf{Prefill queueing time and load balance on CC.}}
\label{tab:app-queue}
\begin{tabular}{@{}lcccc@{}}
\toprule
 & \multicolumn{2}{c}{Dynamo} & \multicolumn{2}{c}{\sys{}} \\
\cmidrule(lr){2-3} \cmidrule(l){4-5}
Concurrent & Queueing & Most loaded / & Queueing & Most loaded / \\
agents & time (s) & average & time (s) & average \\
\midrule
90 & 11.4 & 2.29 & 1.5 & 1.79 \\
108 & 11.0 & 2.38 & 2.2 & 1.75 \\
126 & 11.2 & 2.61 & 4.3 & 1.85 \\
\bottomrule
\end{tabular}
\end{table}
 
\subsection{Throughput and End-to-End Latency}
\label{app:exp-throughput}
 
The stationary runs serve a fixed number of agents in a closed loop
(\autoref{sec:eval-setup}). A branch issues its next request only after its
previous request completes and the pause between the two requests ends, during
which the branch waits for tools, sub-agents, or users. By Little's
law~\cite{little1961proof}, the throughput $X$ of a system satisfies
\begin{equation}
  X = \frac{\bar{N}_{\mathrm{branch}}}
           {\bar{T}_{\mathrm{E2E}} + \bar{T}_{\mathrm{pause}}},
  \label{eq:app-little}
\end{equation}
where $\bar{N}_{\mathrm{branch}}$ is the mean number of requests in flight plus
the mean number of branches in a pause, $\bar{T}_{\mathrm{E2E}}$ is the mean E2E
latency, and $\bar{T}_{\mathrm{pause}}$ is the total pause time divided by the
number of requests. In each setting, the two systems keep a similar
$\bar{N}_{\mathrm{branch}}$, for example 159 and 161 on CC with 126 concurrent
agents. The lower $\bar{T}_{\mathrm{E2E}}$ of \sys{} therefore translates into a
higher throughput (\autoref{tab:app-throughput}).
 
\begin{table}[ht]
\centering
\small
\caption{\textbf{Throughput, mean E2E latency, and pause between
requests on stationary workloads.}}
\label{tab:app-throughput}
\begin{tabular}{@{}llcccccc@{}}
\toprule
 & & \multicolumn{3}{c}{Dynamo} & \multicolumn{3}{c}{\sys{}} \\
\cmidrule(lr){3-5} \cmidrule(l){6-8}
 & Concurrent
 & $X$ & $\bar{T}_{\mathrm{E2E}}$ & $\bar{T}_{\mathrm{pause}}$
 & $X$ & $\bar{T}_{\mathrm{E2E}}$ & $\bar{T}_{\mathrm{pause}}$ \\
Workload & agents
 & (req/s) & (s) & (s)
 & (req/s) & (s) & (s) \\
\midrule
Putnam & 224 & 7.72 & 19.1 & 10.1 & 8.81 & 15.5 & 10.0 \\
 & 256 & 8.37 & 20.4 & 10.1 & 9.20 & 17.8 & 10.0 \\
 & 288 & 9.12 & 21.7 & 10.0 & 10.15 & 18.6 & 9.9 \\
\midrule
CC & 90 & 2.48 & 32.2 & 15.4 & 5.40 & 7.9 & 12.9 \\
 & 108 & 2.48 & 37.7 & 17.1 & 5.78 & 9.9 & 13.5 \\
 & 126 & 2.14 & 54.4 & 18.6 & 5.85 & 13.2 & 14.1 \\
\midrule
OSWorld & 128 & 1.93 & 57.4 & 1.0 & 4.30 & 24.5 & 1.4 \\
 & 144 & 1.98 & 66.3 & 1.4 & 4.33 & 28.1 & 1.4 \\
 & 160 & 1.68 & 84.4 & 1.2 & 4.38 & 31.4 & 1.3 \\
\bottomrule
\end{tabular}
\end{table}
 
Within each period, the run with a shifting workload composition
(\autoref{sec:eval-shift}) also serves a fixed number of agents in a closed loop,
so Equation~\eqref{eq:app-little} holds for this run as well. Over the whole run,
\sys{} reduces the mean E2E latency by 2.15$\times$ and 1.33$\times$ relative to
Dynamo and the best static split, respectively (\autoref{fig:eval-shift-e2e}),
and therefore achieves 2.14$\times$ the throughput of Dynamo and 1.28$\times$
that of the best static split (\autoref{tab:app-shift-throughput}). Its
throughput is also the highest in every period.
 
\begin{table}[ht]
\centering
\small
\caption{\textbf{Throughput (req/s) under the shifting workload composition.} The number in parentheses is the number of concurrent agents, 144 Putnam and 72 OSWorld agents in the mixed period. 5P4D is the best static split.}
\label{tab:app-shift-throughput}
\begin{tabular}{@{}lccc@{}}
\toprule
Period & Dynamo & 5P4D & \sys{} \\
\midrule
Putnam (288) & 1.57 & 3.02 & \textbf{4.05} \\
Mixed (144 + 72) & 2.21 & 3.64 & \textbf{3.99} \\
OSWorld (144) & 2.05 & 3.10 & \textbf{4.43} \\
\midrule
Whole run & 1.94 & 3.25 & \textbf{4.16} \\
\bottomrule
\end{tabular}
\end{table}
 
\section{Additional Experiments}
\label{app:experiments}
 
Beyond the settings of \autoref{sec:evaluations}, we evaluate \sys{} under
larger and more diverse deployments. \sys{} keeps its gain on 256 GB300 GPUs
and, in simulation, on 3{,}072 GB300 GPUs (\appref{app:exp-scale}), sustains twice the agent arrival rate of static 4P5D under open-loop arrivals
(\appref{app:exp-open-loop}), and remains effective when prefill and decode use
different parallelism (\appref{app:exp-hetero-parallel}). Moreover, \sys{}
outperforms ThunderAgent, an agent-aware router, since KV capacity does not
limit serving on GB300 nodes (\appref{app:exp-thrashing}), and its asynchronous
KV fetching achieves 1.12$\times$ the throughput of upstream SGLang
(\appref{app:exp-async-fetch}). \appref{app:exp-static-split} explains why its
best static split differs from that under SGLang PD disaggregation.
 
\subsection{Best Static Split on the Symmetric Engines}
\label{app:exp-static-split}
 
The best static split of \autoref{sec:eval-setup} is profiled offline under SGLang PD disaggregation (\autoref{fig:intro}). \sys{} serves requests on its symmetric engines (\autoref{sec:dataplane-switch}), so its best static split may differ. This subsection compares the best static split of \sys{} with that of SGLang on the stationary workloads of \autoref{sec:eval-single}.
 
\paragraph{Experimental Setup.}
We serve Putnam, CC, and OSWorld with 224, 90, and 128 concurrent agents, the lowest concurrency level of each agent workload in \autoref{sec:eval-single}. As in \autoref{sec:eval-setup}, we serve GLM-5.2 for Putnam and CC and Kimi K2.6 for OSWorld on 36 GB300 GPUs. \emph{SGLang} runs in its PD disaggregation mode at its best static split in \autoref{fig:intro-diversity}, with the router and HiCache settings of our baselines (\autoref{sec:eval-setup}). \emph{\sys{}} runs at its best static split, with online phase-split adaptation disabled and its PD split fixed for the whole run. We score minutes 30 to 85 of each run, and each configuration runs once.
 
\begin{table}[ht]
\centering
\small
\caption{\textbf{The best static split differs between SGLang and the symmetric engines of \sys{}.} Each row is one run of a system at its best static split, the split with the highest throughput, on 36 GB300 GPUs, scored from minute 30 to minute 85. The number in parentheses is the number of concurrent agents.}
\label{tab:app-static-split}
\begin{tabular}{@{}llccrrrr@{}}
\toprule
 & & Best static & Throughput & \multicolumn{2}{c}{TTFT (s)} & \multicolumn{2}{c}{E2E latency (s)} \\
\cmidrule(lr){5-6} \cmidrule(l){7-8}
Workload & System & split & (requests/s) & Mean & P95 & Mean & P95 \\
\midrule
Putnam (224) & SGLang & 2P7D & 9.00 & 1.0 & 2.5 & 14.6 & 74.0 \\
 & \sys{} & 2P7D & \textbf{9.20} & \textbf{0.6} & \textbf{1.2} & \textbf{14.0} & \textbf{72.6} \\
\midrule
CC (90) & SGLang & 4P5D & 4.97 & 4.8 & 19.5 & 9.2 & 30.7 \\
 & \sys{} & 3P6D & \textbf{5.44} & \textbf{3.5} & \textbf{9.7} & \textbf{7.8} & \textbf{22.2} \\
\midrule
OSWorld (128) & SGLang & 6P3D & 3.12 & 35.3 & 50.8 & 35.8 & 51.3 \\
 & \sys{} & 8P1D & \textbf{4.33} & \textbf{21.6} & \textbf{35.6} & \textbf{24.2} & \textbf{39.2} \\
\bottomrule
\end{tabular}
\end{table}
 
\paragraph{Results.}
The symmetric engines shift the best static split from 4P5D to 3P6D for CC and from 6P3D to 8P1D for OSWorld (\autoref{tab:app-static-split}). For Putnam, 2P7D is the best static split on both systems. The shift is largest for OSWorld and stems from how the two systems allocate the GPU memory of decode nodes. SGLang reserves GPU memory on a decode node for the KV of each request before its prefill begins. It holds this reservation while the request waits in the queue of a prefill node. At 6P3D, these reservations occupy 91\% of the KV cache of its three decode nodes on average, so removing a decode node would delay prefill. \sys{} instead selects the decode node of a request only when its prefill begins, so queued requests occupy no memory on decode nodes. With a single decode node, the mean E2E latency of OSWorld exceeds its mean TTFT by only 2.6~s (24.2~s versus 21.6~s). \sys{} therefore assigns the remaining eight nodes to prefill. With online phase-split adaptation, \sys{} reaches 8.81, 5.40, and 4.30 requests/s in these settings (\autoref{tab:app-throughput}), 0.98--1.38$\times$ the throughput of the best static split under SGLang.
 
\begin{figure}[t]
  \centering
  \captionsetup[subfigure]{skip=2pt}
  \includegraphics{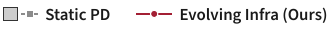}\\[1pt]
  \begin{subfigure}[t]{2.15in}
    \centering
    \captionsetup{margin={0.20in,0in}}
    \includegraphics[width=\linewidth]{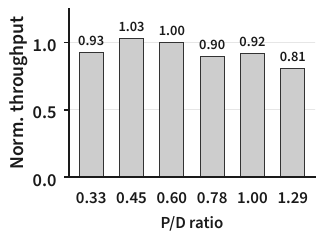}
    \caption{Throughput across P/D ratios}
    \label{fig:app-scale-split}
  \end{subfigure}\hfill
  \begin{subfigure}[t]{2.15in}
    \centering
    \captionsetup{margin={0.20in,0in}}
    \includegraphics[width=\linewidth]{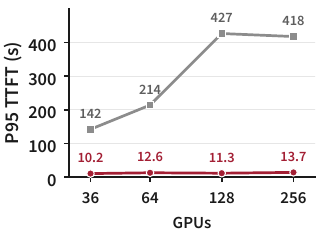}
    \caption{Tail latency at scale}
    \label{fig:app-scale-ttft}
  \end{subfigure}\hfill
  \begin{subfigure}[t]{2.15in}
    \centering
    \captionsetup{margin={0.20in,0in}}
    \includegraphics[width=\linewidth]{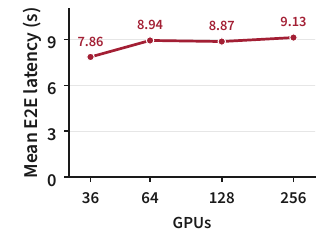}
    \caption{Mean latency at scale}
    \label{fig:app-scale-e2e}
  \end{subfigure}
  \caption{\textbf{\sys{} keeps its latency low as the cluster grows to 256 GPUs.} Static PD is SGLang in its PD disaggregation mode with a fixed split. (a)~On 256 GPUs, throughput is normalized to that at a P/D ratio of 0.6 (24P40D). (b)~From 36 to 256 GPUs, the P95 TTFT of static PD rises about threefold, whereas that of \sys{} stays within 10--14~s. (c)~The mean E2E latency of \sys{} rises by only 16\%.}
  \label{fig:app-scale}
\end{figure}
 
\begin{figure*}[t]
  \centering
  \includegraphics[width=\textwidth]{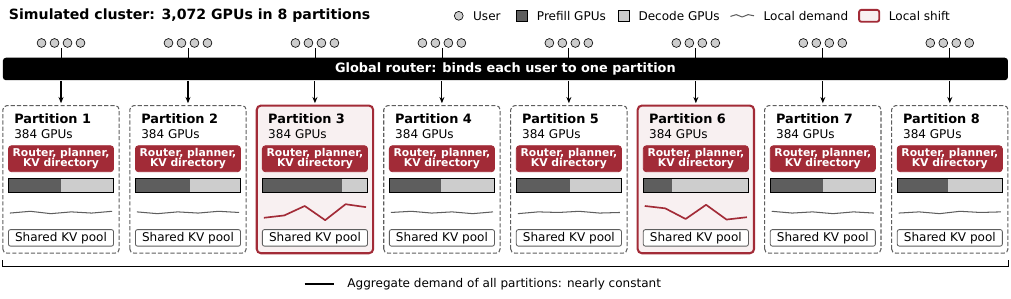}
  \caption{\textbf{Hierarchical serving binds each user to one partition.} In
  a simulated cluster of 3072 GPUs, a global router binds each user to one
  of eight partitions of 384 GPUs. Each partition runs its own router,
  planner, and KV directory. The aggregate demand stays nearly constant, but
  a few partitions (red) see a local shift in demand, and their planners
  adapt their PD splits.}
  \label{fig:app-hierarchy}
\end{figure*}
 
\subsection{Scaling to Larger Clusters}
\label{app:exp-scale}
 
This subsection evaluates \sys{} on real clusters of up to 256 GB300 GPUs
and on a simulated cluster of 3{,}072 GPUs. Both show that the aggregate PD
demand varies less as the cluster grows, whereas the imbalance within the
cluster grows.
 
\paragraph{Setup on up to 256 GPUs.} We serve the Claude Code traces with
GLM-5.2 on 36, 64, 128, and 256 GB300 GPUs and scale the number of
concurrent agents with the GPUs, 10 agents per GPU. Static PD runs
SGLang in its PD disaggregation mode at a fixed split, with the router and
HiCache settings of \autoref{sec:eval-setup}.
 
\paragraph{The aggregate split matters less at scale.} On 256 GPUs, static
PD keeps 81--103\% of the throughput of 24P40D across P/D ratios from 0.33
to 1.29 (\autoref{fig:app-scale}a). The demand of many agents together
varies less over time, and each node holds a smaller fraction of the
capacity. A mismatch between the split and the aggregate demand thus costs
less throughput than on 36 GPUs.
 
\paragraph{The imbalance among nodes grows.} The P95 TTFT of static PD,
however, rises about threefold from 36 to 256 GPUs, from 142~s to 418~s
(\autoref{fig:app-scale}b). With more prefill nodes, the cache-aware router
concentrates the requests of more agents on the few nodes that hold their
prefixes. The router and the shared KV pool of \sys{} spread this load
across nodes. Its P95 TTFT thus stays within 10--14~s, and its mean E2E
latency rises by only 16\% (\autoref{fig:app-scale}c).
 
\paragraph{Offline profiling grows costly at scale.} Finding the best static
split offline also becomes expensive as the cluster grows. A cluster of $N$
nodes with $g$ GPUs each has $N-1$ candidate splits, and evaluating one
candidate replays the whole trace of length $T$ on all $Ng$ GPUs, together
with a warmup of length $T_{\mathrm{w}}$. An exhaustive search thus costs
$(N-1)(T+T_{\mathrm{w}})Ng$ GPU-hours, or $O(N^2 g T)$. A binary search
compares two adjacent splits at each of its $\lceil\log_2(N-1)\rceil$ steps
and costs $O(N g T \log N)$. For the three-hour CC trace on 256 GPUs, $N=64$
and $T_{\mathrm{w}} = 0.5$~h, so each candidate costs 896 GPU-hours. The
binary search takes 12 runs, or 10{,}752 GPU-hours, which is 14$\times$ the
768 GPU-hours of serving the trace once. The exhaustive search takes
56{,}448 GPU-hours. Three factors raise these costs further. First, a binary
search assumes that throughput first rises and then falls with the number
of prefill nodes, yet the throughput in \autoref{fig:app-scale}a rises again
from a P/D ratio of 0.78 to 1.00. A binary search can thus stop at a local
optimum, and a reliable search falls back to the exhaustive one. Second,
adjacent splits differ by only a few percent, for example 3\% between P/D
ratios of 0.45 and 0.60, which is comparable to the variation between runs.
Each candidate then needs repeated runs, which multiplies the cost. Third,
the search must recur for every new workload composition. A best split per
planning window costs even more, since the searches of different windows
visit different splits and together approach all $N-1$ candidates. \sys{}
instead evaluates the $N-1$ candidates with its iteration-time models on $m$
slices at the end of each window, which takes $O(Nm)$ arithmetic on a CPU and
no extra GPU time.
 
\paragraph{Simulation setup.} We further simulate a cluster of 3{,}072 GB300
GPUs with Frontier~\cite{feng2026frontiercomprehensiveaccuratellm}, a
discrete-event simulator for LLM serving. We profile the prefill and decode
iterations of GLM-5.2 offline on our GB300 nodes and supply these profiles
to the simulator. Frontier models KV caching only within GPU memory, and we
extend it with the shared host-memory KV pool and the asynchronous fetches
of \autoref{sec:dataplane-pool}. Since \sys{} fetches remote KV while a
request waits in the queue, the simulator assumes that each fetch completes
before the prefill of its request begins and charges the fetch no time. We
replay the Claude Code and Codex traces over 24 hours with about 10
concurrent agents per GPU, or 30{,}720 agents in total.
 
\paragraph{Hierarchical scheduling.} A single router cannot monitor every
node of a cluster with thousands of GPUs. We therefore divide the cluster
into 8 partitions of 384 GPUs, each with its own router, planner, and KV
directory (\autoref{fig:app-hierarchy}). Constraints such as region and
hardware bind each user to one partition, and each partition thus serves
its own subset of agents.
 
\paragraph{Local shifts under a stable aggregate.} Across all partitions,
the aggregate workload stays nearly constant, and the PD ratio of the
cluster varies by only 7\% over the trace. A static split, however, leaves
2 of the 8 partitions severely imbalanced, with queueing times up to
13.9$\times$ the cluster average. These imbalances rarely stem from bursts
of load. While a partition is imbalanced, its total work changes by less
than 12.3\%, whereas its PD ratio shifts by up to 42.4\%. The planner of
each partition follows these shifts and improves the throughput by
1.42$\times$ over a static split.
 
\paragraph{Summary.} At scale, \sys{} corrects the imbalance at two levels.
Its router and shared KV pool balance the load among the nodes of a
cluster, and the planner of each partition follows the local shifts in PD
demand, even when the aggregate workload of the cluster is fixed. Neither
level requires offline profiling.
 
\subsection{Open-Loop Replay at Fixed Agent Arrival Rates}
\label{app:exp-open-loop}
 
The stationary runs of \autoref{sec:eval-single} replay a fixed number of agents in a closed loop, so a slower system also receives requests more slowly, and its queues stay bounded.
In production, new agents arrive whether or not earlier ones have finished. This subsection lets agents arrive at a fixed rate and measures the highest rate that each system sustains within latency targets.
 
\paragraph{Experimental Setup.}
We serve the Claude Code trace with GLM-5.2 on 36 GB300 GPUs with the models and hardware of \autoref{sec:eval-setup}. New agents arrive in a Poisson process at 0.5 to 1.5 agents per minute in steps of 0.25, and all systems replay the same arrivals at each rate. Within an agent, requests follow the trace as in the stationary runs: each is sent once the request it depends on completes and the recorded delay elapses. At most 258 agents run at once, and the time that an agent waits for admission counts toward the TTFT and E2E latency of its first request. A system sustains a rate if its P50 and P95 TTFT stay within 5~s and 15~s, its P50 E2E latency stays within 10~s, it admits at least 90\% of the arriving agents, and fewer than 0.5\% of its requests fail. We compare \sys{}, which starts from 4P5D, with two baselines of \autoref{sec:eval-setup}: SGLang PD disaggregation at 4P5D, the best closed-loop static split for CC (\autoref{fig:motiv-compute}), and Dynamo, which starts from 6P3D. In the closed-loop runs of \autoref{sec:eval-ablation}, the initial split changes the mean E2E latency of \sys{} by at most 1.8\%. We score minutes 30 to 90 of each run, and each configuration runs once.
 
\begin{table}[ht]
\centering
\small
\caption{\textbf{\sys{} sustains twice the agent arrival rate of static 4P5D.} Each system serves CC once at each arrival rate, and latencies are in seconds. Bold marks the runs that meet every target.}
\label{tab:app-open-loop}
\begin{tabular}{@{}cccccccccc@{}}
\toprule
 & \multicolumn{3}{c}{\sys{}} & \multicolumn{3}{c}{Static 4P5D} & \multicolumn{3}{c}{Dynamo} \\
\cmidrule(lr){2-4} \cmidrule(lr){5-7} \cmidrule(l){8-10}
Arrival rate & \multicolumn{2}{c}{TTFT} & E2E & \multicolumn{2}{c}{TTFT} & E2E & \multicolumn{2}{c}{TTFT} & E2E \\
(agents/min) & P50 & P95 & P50 & P50 & P95 & P50 & P50 & P95 & P50 \\
\midrule
0.5 & \textbf{0.7} & \textbf{6.8} & \textbf{2.5} & \textbf{0.7} & \textbf{10.8} & \textbf{2.7} & 1.0 & 22.0 & 3.4 \\
0.75 & 1.3 & 17.8 & 4.4 & 3.8 & 57.0 & 7.3 & 21.0 & 234.5 & 25.7 \\
1.0 & \textbf{1.2} & \textbf{12.3} & \textbf{4.6} & 6.6 & 139.0 & 11.3 & 11.9 & 122.1 & 17.4 \\
1.25 & 16.8 & 55.0 & 21.8 & 18.5 & 1,142.7 & 21.1 & 16.2 & 186.9 & 21.0 \\
1.5 & 68.4 & 151.9 & 74.7 & 637.8 & 1,377.1 & 639.9 & 19.6 & 251.1 & 24.4 \\
\bottomrule
\end{tabular}
\end{table}
 
\paragraph{Results.}
\sys{} sustains 1.0 agents per minute, twice the rate of static 4P5D, whereas Dynamo misses the targets at every rate, with a P95 TTFT of 22.0~s already at 0.5 agents per minute (\autoref{tab:app-open-loop}). At 1.0 agents per minute, \sys{} keeps the P95 TTFT at 12.3~s, 11.3$\times$ and 9.9$\times$ lower than that of static 4P5D and Dynamo, respectively. As the arrival rate grows, its planner gives prefill more nodes: 2--3 at 0.5 agents per minute, 3--4 at 1.0, 5 at 1.25, and 6--7 at 1.5. \sys{} misses the targets at 0.75 agents per minute, although it meets them at 1.0, because of a single planner switch: at minute 75, the planner moves to 2P7D, and prefill requests queue for 120--275~s at P95 until it adds prefill nodes back at minute 85. Before this switch, the run meets every target, with a P95 TTFT of 14.0~s instead of 17.8~s.
 
\subsection{Compatibility with Heterogeneous Parallelism}
\label{app:exp-hetero-parallel}
 
The best parallelism can differ between prefill and decode. For example, Huawei's xDeepServe serves DeepSeek with tensor parallelism (TP) in prefill and attention data parallelism (ADP) in decode, because the two phases see different sequence lengths and ADP avoids duplicating the MLA KV cache~\cite{xiao2025huawei}. SGLang and vLLM, however, choose parallelism strategy with launch arguments~\cite{sglang2026serverargs,vllm2026dp}, so a node that switches between them must relaunch its engine. The symmetric engine of \sys{} can avoid this relaunch, because where a node places its weights can be decoupled from how it computes with them: a node can hold the union of the weights that both phases need and switch phases by changing only its computation path, without moving any weights. The extra weights take a small share of GPU memory, and the KV cache of \sys{} lies mostly in abundant host memory, which holds four times as much KV as GPU memory (\autoref{sec:eval-setup}).
 
\paragraph{Experimental Setup.}
We serve GLM-5.2~\cite{glm5team2026glm5vibecodingagentic} on six 8$\times$B200 nodes, with TP8 on prefill nodes and ADP8 on decode nodes, and run BrowseComp agents~\cite{wei2025browsecomp}, which can only spawn sub-agents, for two hours. We compare \sys{} with Dynamo~\cite{nvidia2025dynamo} and the best static split, and \sys{} and Dynamo both start from 4P2D. In this deployment, the two phases differ only in the attention layers, and the MoE layers use the same parallelism in both. Under ADP, every GPU holds all attention weights, so it also holds the slice that TP assigns to it. Each node of \sys{} therefore keeps the ADP layout of the attention weights in both phases: it computes attention with TP in prefill and with ADP in decode, and a switch changes only which of the two it runs. With FP8 attention weights, this layout takes about 10~GB more per GPU than TP alone, or 6\% of the memory of a B200 GPU. The planner needs no change, because it fits the iteration-time coefficients of each node online, and these coefficients already depend on the parallelism of the node (\autoref{sec:design-model}).
 
\paragraph{Results.}
As \autoref{tab:app-hetero-parallel} shows, \sys{} moves to 2P4D, the best static split, and reaches a mean E2E latency of 15.9~s, within 1.3\% of the 15.7~s of that split, which requires profiling each split offline. Switching engines takes \sys{} 27~s in total over the two hours. Dynamo ends at 3P3D with a mean E2E latency of 21.9~s, 38\% higher than that of \sys{}. It spends 913~s, 34$\times$ as long, switching engines, because it relaunches the worker of a node at every phase switch (\autoref{sec:motiv-switch}).
 
\begin{table}[ht]
\centering
\small
\caption{\textbf{Phase switches with TP8 prefill and ADP8 decode.} Dynamo and \sys{} start from 4P2D, and the best static split stays fixed. Engine-switch time sums the time of all phase switches over the two hours.}
\label{tab:app-hetero-parallel}
\begin{tabular}{@{}lccc@{}}
\toprule
 & Best static & Dynamo & \sys{} \\
\midrule
Final split & 2P4D & 3P3D & 2P4D \\
Mean E2E latency (s) & 15.7 & 21.9 & 15.9 \\
Engine-switch time (s) & 0 & 913 & 27 \\
\bottomrule
\end{tabular}
\end{table}
 
\paragraph{Wide expert parallelism.}
The same decoupling carries over to wide EP, where one engine spans an EP group of several nodes and attention runs with ADP. Under wide EP, the two phases use groups of different sizes and different communication kernels. SGLang, for example, evaluates prefill on 4-node groups with kernels built for throughput, and decode on 9-node groups with low-latency kernels captured in CUDA graphs~\cite{sglang2025deploying}. A phase switch therefore moves a node from a group of one phase to a group of the other. Elastic EP performs such a move without a restart~\cite{alroy2026elastic}, but it transfers weights: when vLLM grows a DeepSeek-V4-Pro group from 16 to 20 GB200 GPUs, the scale-up takes 92~s in total to transfer weights, warm up, and capture CUDA graphs~\cite{alroy2026achieving}. \sys{} avoids most of this weight transfer by again holding the union of what both groups need. Every node already holds the non-expert weights. For the experts, each group keeps two copies of every expert on different nodes, so each GPU holds twice its share. A node can then leave a group without taking away the last copy of any expert, and it can join any group at least half as large as its old one by serving experts that it already holds. For GLM-5.2 with NVFP4 experts, the second copy takes 7\% of B200 memory in a 4-node group and 3\% in a 9-node group. Nor does a move reload kernels, because one engine can load the kernels of both phases, as the auto mode of SGLang does~\cite{sglang2025deploying}.
 
\subsection{Comparison with Agent-aware Routers}
\label{app:exp-thrashing}
 
ThunderAgent~\cite{kang2026thunderagent} pauses agents to prevent KV-cache
thrashing (\appref{app:related-agent}). \autoref{sec:motiv-kv} instead
attributes the KV reuse that existing systems lose to where they place
requests and which KV they evict. This subsection tests which explanation
holds on GB300 nodes by comparing ThunderAgent with cache-aware routing and
\sys{}.
 
\paragraph{Experimental Setup.}
We serve the Claude Code trace with 126 concurrent agents on 36 GB300 GPUs
with GLM-5.2, which uses DSA (\appref{app:theory-attention}). All three
configurations keep the split at 4P5D (\autoref{tab:app-thrashing}). The
first two run the SGLang engines of our baselines (\autoref{sec:eval-setup}),
which evict host memory in LRU order, and differ only in how they choose the
prefill node. Each of their prefill nodes caches up to 2.06 million tokens of
KV in GPU memory and 8.25 million tokens in host memory, which holds one copy
per node (\appref{app:cpu-dedup}). For \emph{cache-aware} routing, we
reimplement the policy of the SGLang Model Gateway~\cite{sglang2026sglang}.
For \emph{ThunderAgent}, we port the program scheduler that Dynamo ships as a
router plugin~\cite{kang2026thunderagent,nvidia2025dynamo} and keep its
default thresholds. Since the plugin targets aggregated serving, we apply it
to the choice of the prefill node and count host memory in the capacity of a
node. ThunderAgent pauses agents that wait for tools once the contexts of the
active agents on a node exceed 95\% of this capacity. The harness marks the
last request of each agent, so ThunderAgent releases finished agents at once.
The \sys{} configuration is the run with a fixed split in
\autoref{sec:eval-ablation}. We score minutes 30 to 85 of each run, and each
configuration runs once.
 
\begin{table}[t]
\centering
\small
\caption{\textbf{KV capacity does not limit serving on GB300 nodes.} Each
column is one run of the Claude Code trace with 126 concurrent agents on
GLM-5.2, 36 GB300 GPUs, and a fixed 4P5D split, scored from minute 30 to
minute 85. The ideal KV hit rate counts every prompt token whose KV an earlier
request produced. Lost KV reuse is the share of these tokens that a run
recomputes.}
\label{tab:app-thrashing}
\begin{tabular}{@{}lccc@{}}
\toprule
 & Cache-aware & ThunderAgent & \sys{} \\
\midrule
Throughput (requests/s) & 4.66 & 4.43 & \textbf{6.27} \\
TTFT, mean / P95 (s) & 15.3 / 60.3 & 17.5 / 59.8 & \textbf{4.5} / \textbf{14.0} \\
E2E latency, mean / P95 (s) & 19.7 / 66.7 & 22.0 / 67.1 & \textbf{11.2} / \textbf{32.6} \\
TPOT, mean (ms) & 4.6 & 4.4 & 7.9 \\
\midrule
KV hit rate, actual / ideal (\%) & 94.9 / 97.2 & 94.7 / 97.2 & 96.6 / 97.3 \\
Lost KV reuse (\%) & 2.4 & 2.6 & \textbf{0.7} \\
Uncached prompt tokens per request & 7,660 & 8,119 & \textbf{5,268} \\
Peak KV demand of active agents / capacity & -- & 0.86 & -- \\
Paused requests (\%) & -- & 0 & -- \\
\bottomrule
\end{tabular}
\end{table}
 
\paragraph{Results.}
ThunderAgent pauses no request (\autoref{tab:app-thrashing}): by its own
estimate, the active agents on a prefill node need at most 86\% of its KV
capacity (76\% on average), below its 95\% threshold, since the deduplicated
host tier absorbs thrashing (\appref{app:cpu-dedup}). It thus performs like
cache-aware routing, with 2.6\% versus 2.4\% lost KV reuse, 0.95$\times$ the
throughput, and 1.14$\times$ the mean TTFT. With the same split and
host-memory ratio, \sys{} achieves \textbf{1.35$\times$} the throughput and
0.29$\times$ the mean TTFT of cache-aware routing and loses only 0.7\% of the
KV reuse, because its prefill nodes fetch KV from any node that holds it
(\autoref{sec:dataplane-pool}) and evict first the copies that no active
branch references (\autoref{sec:dataplane-directory}). Its higher TPOT
coincides with larger decode batches, and this comparison does not separate
the effects of routing and the data plane. On GB300 nodes, KV reuse therefore
depends on where KV resides and which copies host memory keeps, not on KV
capacity (\autoref{sec:motiv-kv}).
 
\subsection{Async KV Fetching}
\label{app:exp-async-fetch}
 
When a prefill node fetches KV from host memory, it restores the KV into GPU
memory and broadcasts it to all GPUs of the node
(\autoref{sec:dataplane-pool}). This subsection compares the load-back of
upstream SGLang with the asynchronous KV fetching of \sys{}.
 
\paragraph{Experimental Setup.}
We use the setup of \appref{app:exp-thrashing}: the Claude Code trace with
126 concurrent agents, GLM-5.2 on 36 GB300 GPUs, a fixed 4P5D split, and
deduplicated host memory (\appref{app:cpu-dedup}). The router uses
cache-aware routing for prefill nodes and round-robin for decode nodes in
both runs, and only the engine data plane differs. \emph{Upstream SGLang}
broadcasts the KV to the GPUs of the node with NCCL collectives, which share
the GPUs with the prefill computation. \sys{} moves the KV between GPUs with
their copy engines over NVLink while the request waits in the queue, and its
decode nodes write KV through to host memory with a GPU kernel that does not
block the scheduler. We score minutes 30 to 85 of each run, and each
configuration runs once.
 
\begin{table}[t]
\centering
\small
\caption{\textbf{Asynchronous KV load-back.} Each column is one run of the
Claude Code trace with 126 concurrent agents on GLM-5.2, 36 GB300 GPUs, and a
fixed 4P5D split with the same routing, scored from minute 30 to minute 85.
Idle GPU time counts the time between iterations while a node has requests
to serve.}
\label{tab:app-async-fetch}
\begin{tabular}{@{}lcc@{}}
\toprule
 & Upstream SGLang & \sys{} \\
\midrule
Throughput (requests/s) & 4.62 & \textbf{5.18} \\
TTFT, mean / P95 (s) & 15.7 / 56.4 & \textbf{12.1} / \textbf{42.8} \\
E2E latency, mean / P95 (s) & 20.2 / 64.5 & \textbf{16.5} / \textbf{50.7} \\
TPOT, mean / P95 (ms) & 4.4 / 8.1 & 4.5 / 8.2 \\
\midrule
Idle GPU time, prefill / decode (\%) & 23.6 / 9.0 & \textbf{14.4} / \textbf{7.8} \\
KV hit rate, actual / ideal (\%) & 95.1 / 97.3 & 94.9 / 97.3 \\
Uncached prompt tokens/s (thousands) & 34.3 & 39.3 \\
\bottomrule
\end{tabular}
\end{table}
 
\paragraph{Results.}
With the same routing, the asynchronous load-back of \sys{} cuts the idle
GPU time of prefill nodes from 23.6\% to 14.4\% and achieves
\textbf{1.12$\times$} the throughput of upstream SGLang
(\autoref{tab:app-async-fetch}). Mean and P95 TTFT drop by 23\% and 24\%,
and E2E latency by 18\% and 21\%, although \sys{} serves 15\% more uncached
prompt tokens per second. TPOT stays within 3\%.
In upstream SGLang, the fetch delays prefill substantially. A prefill node
must load the fetched KV into GPU memory before the layers that read it can
run, and the NCCL broadcast of this KV shares the GPUs with the prefill
computation. As a result, 93\% of its prefill iterations wait more than 10~ms
before they start, and waiting for the load-back alone takes up to 8\% of the
prefill GPU time, about 0.3~s for each iteration that consumes fetched KV.
These waits add to the queueing delay of every request that fetches its
prefix (\autoref{sec:motiv-kv}).
 
\paragraph{Fetches on decode nodes.} The same runs show why the decode nodes
of \sys{} only write to the pool (\autoref{sec:dataplane-pool}). In upstream
SGLang, a decode node also loads reusable KV from its host memory, and the
whole batch of the node waits for the load. Decode iterations that start
while a load-back is in flight leave the GPUs idle 58.7\% of the time, with a
mean gap of 82~ms before the forward, about 19 times the mean TPOT of 4.4~ms,
against 6.9\% for the other decode iterations. Such iterations are rare
(0.5\% of decode iterations), but each one stalls every request on the node.
 
\begin{figure}[!htb]
\centering
\includegraphics{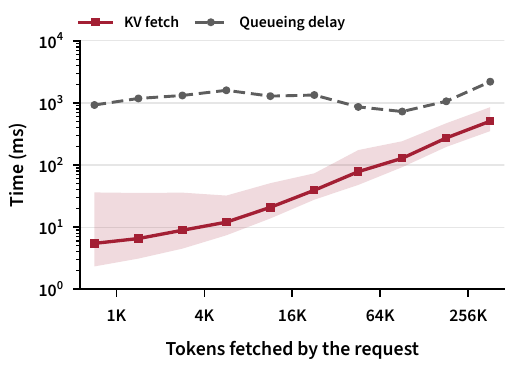}
\caption{\textbf{Time of KV fetches between prefill nodes.} Same configuration
as the run without online phase-split adaptation in
\autoref{fig:eval-ablation-adaptation}. Lines show medians over power-of-two
bins of fetched tokens, and the band shows the 10th to 90th percentiles of the
fetch time. The fetch time grows almost linearly with the fetched tokens and
stays well below the queueing delay, i.e., the time from the arrival of a
request at its prefill node to the start of its prefill.}
\label{fig:app-fetch-latency}
\end{figure}
 
\paragraph{Fetches between prefill nodes.} A prefill node may receive a
request whose prefix KV resides in the host memory of another prefill node.
The node then fetches this KV into its own host memory over RDMA
(\autoref{sec:dataplane-pool}). We profile these fetches in the run
without online phase-split adaptation (\autoref{fig:eval-ablation-adaptation})
from minute 30 to minute 85. As \autoref{fig:app-fetch-latency} shows, the
fetch time grows almost linearly with the number of fetched tokens. Each token
carries 54.6~KB of KV, and the RDMA transfer reaches about 52~GB/s for large
fetches. A fetch of 64K--128K tokens therefore takes 129~ms at the median.
Most fetches are much smaller. For the 92\% of fetches with fewer than 16K
tokens, the median fetch time is 7~ms. \sys{} starts the fetch as soon as the request arrives at the prefill node that serves it. The fetch therefore overlaps with the \emph{queueing delay} of the
request, i.e., the time from this arrival to the start of its prefill. The
median queueing delay is 0.7--2.2~s, at least 3.9 times the median fetch time
for every fetch size. A fetch that starts only when the prefill begins would
instead add its full time to the TTFT.
 
\section{Additional Details}
\label{app:details}
 
This section gives the three algorithms of \sys{} in pseudocode.
\autoref{alg:branch-version} and \autoref{alg:refcount-evict} implement the
branch counts of the cluster-wide KV directory
(\autoref{sec:dataplane-directory}) and the host-memory eviction of
\autoref{sec:design-placement}. \autoref{alg:planner} combines the routing
of \autoref{sec:design-placement} (inner loop) with the phase-split planner
of \autoref{sec:design-split} (outer loop).
 
\subsection{Branch and version identifiers}
\label{app:details-branch}
 
A branch is an independently maintained context and its sequence of requests
(\autoref{sec:introduction}). The harness tags each request with the ID
$\beta$ of the branch that issues it. When it forks a context or spawns a
sub-agent, it uses a new ID for the new branch, and it reports when a branch
finishes. The router keeps, for each branch, its version $v_\beta$, the
prefix-tree path $\pi_\beta$ of its last request, and the nodes $N_\beta$ where
it holds counts. A path is the sequence of KV blocks of a context in the
prefix tree of the KV directory. The version tells an appended context from a
rewritten one (\autoref{alg:branch-version}): a request that only appends
tokens extends the previous path and keeps the version, whereas a request
whose context was rewritten, e.g.\ by compaction, leaves the previous path
and increments the version.
 
\begin{algorithm}[h]
\caption{Branch ID and version of a request}
\label{alg:branch-version}
\renewcommand{\algorithmicrequire}{\textbf{Input:}}
\renewcommand{\algorithmicensure}{\textbf{Output:}}
\begin{algorithmic}[1]
\REQUIRE request $q$ with branch ID $\beta$ from the harness; prefix tree $\mathcal{T}$ of the KV directory; branch table $S$
\ENSURE version $v_\beta$, previous path $\pi_{\mathrm{old}}$, new path $\pi_{\mathrm{new}}$
\STATE $\pi_{\mathrm{new}} \leftarrow$ path of the full pages of the input of $q$ in $\mathcal{T}$, including the blocks that the prefill of $q$ adds
\IF{$\beta \notin S$}
  \STATE $S[\beta] \leftarrow (v_\beta{=}1,\ \pi_\beta{=}\emptyset,\ C_{\beta,n}{=}\emptyset \text{ for every node } n)$ \COMMENT{new, forked or sub-agent branch}
  \STATE $\pi_{\mathrm{old}} \leftarrow \emptyset$
\ELSE
  \STATE $\pi_{\mathrm{old}} \leftarrow \pi_\beta$
  \IF{$\pi_{\mathrm{old}}$ is not a prefix of $\pi_{\mathrm{new}}$}
    \STATE $v_\beta \leftarrow v_\beta + 1$ \COMMENT{context rewritten, e.g.\ by compaction}
  \ENDIF
\ENDIF
\STATE $\pi_\beta \leftarrow \pi_{\mathrm{new}}$
\STATE \textbf{return} $(v_\beta, \pi_{\mathrm{old}}, \pi_{\mathrm{new}})$
\end{algorithmic}
\end{algorithm}
 
A forked branch starts with a new ID and an empty previous path, so the node
that serves it adds one count to every block that it reuses from its parent.
A sub-agent starts a new context, so its blocks are new to the tree. Neither
case changes the counts of the parent branch.
 
\subsection{Branch counts and host-memory eviction}
\label{app:details-refcount}
 
The directory keeps, for each block $b$ and each node $n$ that holds it, the
count $R_{b,n}$ of active branches that reference $b$ on $n$.
\autoref{alg:refcount-evict} updates these counts from the output of
\autoref{alg:branch-version} and from KV transfers, and each node evicts its
host blocks in the order that the counts give
(\autoref{sec:design-placement}).
 
\begin{algorithm}[h]
\caption{Branch counts in the KV directory and host-KV eviction}
\label{alg:refcount-evict}
\renewcommand{\algorithmicrequire}{\textbf{Input:}}
\renewcommand{\algorithmicensure}{\textbf{Output:}}
\begin{algorithmic}[1]
\REQUIRE counts $R_{b,n}$; branch table $S$ with $v_\beta$, $\pi_\beta$, and
  the counted blocks $C_{\beta,n}$ of each branch on each node; requests,
  placements, KV transfers, branch ends, and phase switches
\ENSURE updated counts; host blocks to evict on each node
\STATE \textbf{Count}$(\beta, n, \pi)$: \COMMENT{make $\beta$ count exactly the blocks of $\pi$ on $n$}
\STATE \quad $R_{b,n} \leftarrow R_{b,n} + 1$ for $b \in \pi \setminus C_{\beta,n}$; \
  $R_{b,n} \leftarrow R_{b,n} - 1$ for $b \in C_{\beta,n} \setminus \pi$
\STATE \quad $C_{\beta,n} \leftarrow \pi$; \ $N_\beta \leftarrow \{m : C_{\beta,m} \neq \emptyset\}$
\STATE \textbf{Release}$(\beta, n)$: Count$(\beta, n, \emptyset)$
\STATE \textbf{on} request of branch $\beta$ with $(v_\beta, \pi_{\mathrm{old}}, \pi_{\mathrm{new}})$ from \autoref{alg:branch-version}:
\IF{$v_\beta$ increased}
  \STATE Count$(\beta, m, C_{\beta,m} \cap \pi_{\mathrm{new}})$ for $m \in N_\beta$ \COMMENT{release the dropped history once per node}
\ENDIF
\STATE \textbf{on} placing $\beta$ on node $n$ (prefill routing or decode binding):
\STATE \quad Count$(\beta, n, \pi_\beta)$ \COMMENT{new blocks $+1$; reused blocks keep their counts}
\STATE \quad \textbf{for all} $m \in N_\beta \setminus \{n\}$ in the phase of $n$: once $n$ holds
  version $v_\beta$ of $\beta$, Release$(\beta, m)$ \COMMENT{P2P or D2D move}
\STATE \quad nodes of the other phase keep their counts \COMMENT{P2D or D2P: both copies serve later steps}
\STATE \textbf{on} branch end of $\beta$: Release$(\beta, m)$ for $m \in N_\beta$; remove $\beta$ from $S$
\STATE \textbf{on} phase switch of $n$ from D to P: Release$(\beta, n)$ for every $\beta$ with $n \in N_\beta$
  \COMMENT{reset the counts on $n$}
\STATE \textbf{on} phase switch of $n$ from P to D: keep the counts on $n$
\STATE \textbf{EvictHost}$(n, s)$: $\mathcal{H} \leftarrow$ unpinned blocks in the host memory of $n$
\WHILE{less than $s$ bytes freed \AND $\mathcal{H} \neq \emptyset$}
  \STATE $b^{*} \leftarrow \arg\min_{b \in \mathcal{H}} \bigl(\mathds{1}[R_{b,n} > 0],\ T_b\bigr)$
    \COMMENT{zero counts first, LRU within each group}
  \STATE free the host copy of $b^{*}$; \ $\mathcal{H} \leftarrow \mathcal{H} \setminus \{b^{*}\}$
  \STATE remove $b^{*}$ from $C_{\beta,n}$ for every $\beta$; \ $R_{b^{*},n} \leftarrow 0$
\ENDWHILE
\end{algorithmic}
\end{algorithm}
 
In \autoref{alg:refcount-evict}, $T_b$ is the last access time of block $b$
and the minimum follows lexicographic order. A block that an active branch
still references on $n$ is evicted only after no zero-count block remains,
which keeps the KV of a branch that waits for a tool and evicts the history
that compaction replaced or the source copy of a moved branch.
 
\subsection{Routing and phase-split planning}
\label{app:details-planner}
 
\autoref{alg:planner} runs two loops. The inner loop routes each request to the
prefill node with the shortest predicted TTFT, Equation~\eqref{eq:ttft}. For
each prefill node $n$, the router forms the work of the request on $n$: the
cached prefix $t_{\mathrm{c},n}$ is the longest prefix that the KV directory
finds on $n$ or on another node, of which $x_n$ tokens must be fetched from
another node, and $t_{\mathrm{uc},n}$ counts the remaining input tokens. The
router reads the running batch and the queue of $n$ from the latest state that
$n$ reports, and adds the requests that it has sent to $n$ but that $n$ has not
reported yet. It sums the service times of Equation~\eqref{eq:prefill-model} to
predict the wait $\hat{W}_n$ before the request starts. The fetch runs while the
request waits, so only the part of the fetch time $\tau(x_n)$ that exceeds the
wait adds to the prediction. The prediction ends when the prefill of the request
finishes, so it leaves out the first decode iteration, which is the same on
every prefill node. Our implementation evaluates the sum by replaying the
chunked prefill iterations of $n$ in arrival order, which also accounts for
requests that share an iteration. The decode node is bound when the prefill
starts. The outer loop is the planner of \autoref{sec:design-split}.
 
\begin{algorithm}[h]
\caption{Online adaptation: prefill routing (inner loop) and phase-split planning (outer loop)}
\label{alg:planner}
\renewcommand{\algorithmicrequire}{\textbf{Input:}}
\begin{algorithmic}[1]
\REQUIRE nodes $\mathcal{N}$, initial split $(\mathcal{P},\mathcal{D})$, window $T_{\mathrm{win}}$ with $m$ slices, warmup $T_{\mathrm{learn}}$, fetch time $\tau(\cdot)$, copy price $\kappa$
\STATE Serve $T_{\mathrm{learn}}$ in a collocated mode; fit the iteration-time models of each node; freeze the coefficients; start PD disaggregation from $(\mathcal{P},\mathcal{D})$
\LOOP
  \FORALL{request $r$ that arrives in the window}
    \STATE Update the branch of $r$ with \autoref{alg:branch-version} and \autoref{alg:refcount-evict}
    \FORALL{$n \in \mathcal{P}$}
      \STATE $t_{\mathrm{c},n} \leftarrow$ longest prefix of $r$ cached on $n$ or on another node; \ $x_n \leftarrow$ its tokens that only another node holds
      \STATE $w_n \leftarrow$ prefill part of the work of $r$, with $t_{\mathrm{uc},n} = t_{\mathrm{in}}(r) - t_{\mathrm{c},n}$ and $t_{\mathrm{c},n}$; \ $Q_n \leftarrow$ queue of $n$ plus requests sent to $n$ but not yet reported
      \STATE $\hat{W}_n \leftarrow T^{\mathrm{rem}}_n + \sum_{j \in Q_n} T^{\mathrm{req}}_{\mathrm{P},n}(w_j)$ \COMMENT{wait before $r$ starts}
      \STATE $\hat{T}_n \leftarrow \hat{W}_n + T^{\mathrm{req}}_{\mathrm{P},n}(w_n) + \max\bigl(0,\ \tau(x_n) - \hat{W}_n\bigr) + \kappa\, x_n$ \COMMENT{Eq.~\eqref{eq:ttft} plus the exposed fetch and the copy price}
    \ENDFOR
    \STATE $n^{*} \leftarrow \arg\min_{n \in \mathcal{P}} \hat{T}_n$, ties to the node with the longest prefix of $r$ in its own memory; \ append $r$ to $Q_{n^*}$
    \STATE Send $r$ to $n^{*}$, which fetches the $x_{n^*}$ tokens while $r$ waits; place the branch of $r$ on $n^{*}$ (\autoref{alg:refcount-evict})
    \STATE When $n^{*}$ starts to prefill $r$, bind $r$ to the node in $\mathcal{D}$ with KV room for $r$ and the fewest requests, and place the branch of $r$ on that node
  \ENDFOR
  \STATE Divide the window into $m$ slices; measure $\bar{w}_i, \bar{q}_i$ on each slice $i$; remeasure $\bar{t}_n$ and $f_n$ of each prefill node
  \IF{no node is still draining from the last switch \AND at least 120~s have passed since it}
    \STATE $\mathcal{C} \leftarrow \emptyset$
    \FORALL{prefill count $p \neq |\mathcal{P}|$}
      \STATE $(\mathcal{P}_p, \mathcal{D}_p) \leftarrow$ split with $p$ prefill nodes that changes the fewest nodes
      \STATE $g_i \leftarrow \Theta(\mathcal{P}_p,\mathcal{D}_p;\bar{w}_i,\bar{q}_i) - \Theta(\mathcal{P},\mathcal{D};\bar{w}_i,\bar{q}_i)$ for $i = 1,\dots,m$
      \STATE $\bar{g}_p, \sigma_p \leftarrow$ mean and standard deviation of $g_1,\dots,g_m$
      \IF{$\bar{g}_p > 2\sigma_p/\sqrt{m}$}
        \STATE $\mathcal{C} \leftarrow \mathcal{C} \cup \{p\}$
      \ENDIF
    \ENDFOR
    \IF{$\mathcal{C} \neq \emptyset$}
      \STATE $p^{*} \leftarrow \arg\max_{p \in \mathcal{C}} \bar{g}_p$; apply $(\mathcal{P}_{p^*}, \mathcal{D}_{p^*})$ through the phase switch
    \ENDIF
  \ENDIF
\ENDLOOP
\end{algorithmic}
\end{algorithm}
 
\textbf{Parameters of our runs.} Each prefill node reports its state at every
iteration launch, and the router uses a state only if it is at most 5\,s old.
The chunk size is 8{,}192 tokens. $\tau(x) = 0.45\,\mathrm{s} +
1\,\mu\mathrm{s}\cdot x$ for $x \ge 512$ tokens (no fetch below 512 tokens).
The router also adds to $\hat{T}_n$ a copy price proportional to $x_n$, which
charges the duplicate KV that a fetch creates. $T_{\mathrm{learn}} =
1{,}800$\,s, $T_{\mathrm{win}} = 600$\,s, and 60-s slices, so $m = 10$ for a
full window. The planner skips the 120\,s after each switch.

\end{document}